\documentclass[aps,prl,superscriptaddress,floatfix,twocolumn,notitlepage,nofootinbib]{revtex4-1}

\usepackage{amsmath,amssymb,amsthm}
\usepackage{dsfont}
\usepackage{braket}
\usepackage{graphicx}
\usepackage{comment}
\usepackage{latexsym,stmaryrd}
\usepackage[colorlinks,allcolors=blue]{hyperref}

\usepackage[capitalize]{cleveref}

\newcommand{\ketbra}[1]{|#1\rangle\langle#1|}
\newcommand{\ketbrat}[2]{|#1\rangle\langle#2|}

\DeclareMathOperator{\Tr}{Tr}
\DeclareMathOperator{\poly}{poly}

\DeclareMathOperator{\spn}{span}

\DeclareMathOperator{\rank}{rank}

\newcommand{\Id}{\mathds{1}}
\newcommand{\eps}{\mathcal{E}}

\newcommand{\CC}{\mathbb{C}}

\makeatletter
\newtheoremstyle{mytheorem}
  {3pt}
  {3pt}
  {}
  {}
  {\itshape\bfseries}
  {:}
  {.5em}
  {\thmname{#1}\thmnumber{\@ifnotempty{#1}{ }#2}%
   \thmnote{ {\the\thm@notefont(#3)}}}
\makeatother
\theoremstyle{mytheorem}

\newtheorem{defn}{Definition}
\newtheorem{lemma}{Lemma}
\newtheorem{claim}{Claim}
\newtheorem{prop}{Proposition}
\newtheorem{thm}{Theorem}
\newtheorem{coro}{Corollary}
\newtheorem*{thmCMP}{Theorem CMP18}

\newcommand{\ideal}{\textnormal{ideal}}
\newcommand{\local}{\textnormal{local}}
\newcommand{\NN}{\textnormal{NN}}

\newcommand{\circuit}{\textnormal{circuit}}
\newcommand{\sparse}{\textnormal{sparse}}
\newcommand{\state}{\textnormal{state}}
\newcommand{\rest}{\textnormal{rest}}
\newcommand{\extra}{\textnormal{extra}}
\newcommand{\anc}{\textnormal{anc}}
\newcommand{\eff}{\textnormal{eff}}
\newcommand{\clock}{\textnormal{clock}}
\newcommand{\cS}{\mathcal{S}}
\newcommand{\cN}{\mathcal{N}}

\newcommand{\F}{\mathcal{F}}
\newcommand{\qedextra}{\hfill\ensuremath{\blacklozenge}}

\newcommand{\PE}{\textnormal{PE}}

\newcommand{\A}{\mathcal{A}}
\renewcommand{\L}{\mathcal{L}}
\renewcommand{\H}{\mathcal{H}}

\newcommand{\blnk}{ 
\kern-1bp
\setlength{\unitlength}{10bp}
\begin{picture}(1,1)
\put(0.15,0.05){$\bigcirc$}
\end{picture}
\kern+3bp
}
\newcommand{\lmove}{ 
\setlength{\unitlength}{10bp}
\kern-1bp
\begin{picture}(1,1)
\put(0.27,0.05){$\shortleftarrow$}
\put(0.15,0.05){$\bigcirc$}
\end{picture}
\kern+3bp
}
\newcommand{\turn}{ 
\setlength{\unitlength}{10bp}
\kern-1bp
\begin{picture}(1,1)
\put(0.27,0.05){$\circlearrowleft$}
\put(0.15,0.05){$\bigcirc$}
\end{picture}
\kern+3bp
}
\newcommand{\insi}{ 
\setlength{\unitlength}{10bp}
\kern-1bp
\begin{picture}(1,1)
\put(0.4,0.05){$\circ$}
\put(0.15,0.05){$\bigcirc$}
\end{picture}
\kern+3bp
}
\newcommand{\dead}{ 
\kern-1bp
\setlength{\unitlength}{10bp}
\begin{picture}(1,1)
\put(0.27,0.05){$\times$}
\put(0.15,0.05){$\bigcirc$}
\end{picture}
\kern+3bp
}
\newcommand{\qubit}{ 
\setlength{\unitlength}{10bp}
\kern+1bp
\begin{picture}(1,1)
\linethickness{1bp}
\put(0,-0.2){\line(0,1){1}}
\put(0,-0.2){\line(1,0){1}}
\put(1,0.8){\line(0,-1){1}}
\put(1,0.8){\line(-1,0){1}}
\end{picture}
\kern+1bp
}
\newcommand{\gate}{ 
\setlength{\unitlength}{10bp}
\kern+1bp
\begin{picture}(1,1)
\put(0.1,0){$\blacktriangleright$}
\linethickness{1bp}
\put(0,-0.2){\line(0,1){1}}
\put(0,-0.2){\line(1,0){1}}
\put(1,0.8){\line(0,-1){1}}
\put(1,0.8){\line(-1,0){1}}
\end{picture}
\kern+1bp
}
\newcommand{\rmove}{ 
\setlength{\unitlength}{10bp}
\kern+1bp
\begin{picture}(1,1)
\put(0.1,0){$\vartriangleright$}
\linethickness{1bp}
\put(0,-0.2){\line(0,1){1}}
\put(0,-0.2){\line(1,0){1}}
\put(1,0.8){\line(0,-1){1}}
\put(1,0.8){\line(-1,0){1}}
\end{picture}
\kern+1bp
}
\newcommand{\abet}{ 
\kern+1bp
\setlength{\unitlength}{10bp}
\begin{picture}(1,1)
\put(0.15,-0.05){$\beta$}
\linethickness{1bp}
\put(0,-0.2){\line(0,1){1}}
\put(0,-0.2){\line(1,0){1}}
\put(1,0.8){\line(0,-1){1}}
\put(1,0.8){\line(-1,0){1}}
\end{picture}
\kern+1bp
}
\newcommand{\alpa}{ 
\kern+1bp
\setlength{\unitlength}{10bp}
\begin{picture}(1,1)
\put(0.15,0.05){$\alpha$}
\linethickness{1bp}
\put(0,-0.2){\line(0,1){1}}
\put(0,-0.2){\line(1,0){1}}
\put(1,0.8){\line(0,-1){1}}
\put(1,0.8){\line(-1,0){1}}
\end{picture}
\kern+1bp
}
\newcommand{\lqubit}{ 
\kern+1bp
\setlength{\unitlength}{10bp}
\begin{picture}(1,1)
\put(0.15,0.05){$\scriptstyle{L}$}
\linethickness{1bp}
\put(0,-0.2){\line(0,1){1}}
\put(0,-0.2){\line(1,0){1}}
\put(1,0.8){\line(0,-1){1}}
\put(1,0.8){\line(-1,0){1}}
\end{picture}
\kern+1bp
}
\newcommand{\rqubit}{ 
\kern+1bp
\setlength{\unitlength}{10bp}
\begin{picture}(1,1)
\put(0.15,0.05){$\scriptstyle{R}$}
\linethickness{1bp}
\put(0,-0.2){\line(0,1){1}}
\put(0,-0.2){\line(1,0){1}}
\put(1,0.8){\line(0,-1){1}}
\put(1,0.8){\line(-1,0){1}}
\end{picture}
\kern+1bp
}
\newcommand{\parity}{ 
\setlength{\unitlength}{10bp}
\begin{picture}(0.3,1)
\put(0.15,-0.3){\line(0,1){1.2}}
\end{picture}
}
\newcommand{\bdry}{ 
\setlength{\unitlength}{10bp}
\begin{picture}(0.3,1)
\put(0.07,-0.3){\line(0,1){1.2}}
\put(0.23,-0.3){\line(0,1){1.2}}
\end{picture}
}

\newcommand{\idle}{ 
\kern-1bp
\setlength{\unitlength}{10bp}
\begin{picture}(1,1)
\put(0.27,0.05){$\blacklozenge$}
\put(0.15,0.05){$\bigcirc$}
\end{picture}
\kern+3bp
}

\newcommand{\clkbdry}{ 
\setlength{\unitlength}{10bp}
\begin{picture}(0.6,1)
\put(0.15,-0.3){\line(0,1){1.2}}
\put(0.15,-0.3){\line(1,4){0.3}}
\put(0.45,-0.3){\line(-1,4){0.3}}
\put(0.45,-0.3){\line(0,1){1.2}}
\end{picture}
}

\newcommand{\mgdots}{\cdots}

\begin{document}

\author{Leo Zhou}
\email{leoxzhou@ucla.edu}
\affiliation{\small Electrical and Computer Engineering Department, University of California, Los Angeles, CA 90095, USA}
\affiliation{\small Walter Burke Institute for Theoretical Physics, Caltech, Pasadena, California 91125, USA}
\affiliation{\small Institute for Quantum Information and Matter, Caltech, Pasadena, California 91125, USA}

\author{Dorit Aharonov}
\email{dorit.aharonov@gmail.com}
\affiliation{\small School of Computer Science and Engineering, The Hebrew University, Jerusalem 91904, Israel}

\title{Universal Hamiltonian simulators in one and two dimensions}

\date{August 28, 2026}

\begin{abstract}
A major quantum computing application is analog Hamiltonian simulations, in which the low-lying spectrum of a simulator Hamiltonian $H'$ encodes the physics of a target Hamiltonian $H$.
Some families of 2D spin-lattice Hamiltonians---such as the Heisenberg or XY model on the square lattice---are known to be ``universal'' simulators, in the sense that they can simulate any target Hamiltonian.
Unfortunately, while the known simulations of 1D or 2D target Hamiltonians are efficient, they require resources growing exponentially with system sizes for target Hamiltonians with general connectivity.
This leaves many important cases such as 3D or all-to-all-connected target Hamiltonians out of reach in practice.

Here we remedy this situation and show how the known 2D universal families of Hamiltonians and a new 1D family can simulate all target local Hamiltonians, with only polynomial-scaling overhead. 
This exponential improvement is achieved by a nonperturbative method combining the quantum phase-estimation algorithm and the circuit-to-Hamiltonian construction.
Furthermore, all Hamiltonians efficiently simulable by quantum circuits, including nonlocal ones, also have efficient analog simulators in our 2D and 1D universal families.
Our work establishes that analog simulations of general Hamiltonians can be made efficient, significantly expanding the application potential of analog Hamiltonian simulations in near-term quantum technologies.
\end{abstract}

\maketitle

Analog Hamiltonian simulations enable reproducing the physics of a given target Hamiltonian $H$ by another, easier to implement Hamiltonian $H'$.
Such simulations have a vast range of possible applications, e.g., probing new many-body physics, developing new materials and drugs~\cite{Arguello2019}, as well as implementing Hamiltonian-based quantum algorithms~\cite{FarhiAdiabatic2000, AharonovAQCUniversal, farhi2014quantum}.
Indeed, simulating many-body quantum physics was the major motivation in Feynman's 1981 proposal~\cite{Feynman1982} for quantum computers. 
Cirac and Zoller~\cite{CiracZollerNatPhys2012} further argued for the importance of analog simulations based on their less stringent requirements on error correction and controls compared to fault tolerant quantum computers.
This robustness suggests that such simulations could already be useful in noisy quantum devices~\cite{Preskill2018}.
Exciting demonstrations of this approach have been observed in recent years, where coherent analog Hamiltonian simulations in systems as large as thousands of quantum particles~\cite{Daley2022} have been successfully realized to address many-body physics questions~\cite{Choi2016,Bernien2017,Zhang2017,Kokail2019,Ebadi2020,Periwal2021,Zhang2024observation,Andersen2025thermalization}.

For analog Hamiltonian simulators to be practical for a wide range of applications, an important consideration is that the simulator should be easy to implement, and not be re-engineered on a case-by-case basis.
Furthermore, it is often highly desirable for the simulator to be embeddable in a simple architecture with easy physical realization, e.g., 
a low-dimensional lattice.

An important step towards such a ``universal Hamiltonian simulator" theory was taken by Cubitt, Montanaro, and Piddock in Ref.~\cite{UniversalHamiltonian}. 
There, they took inspiration from the notion of \emph{a universal set of quantum gates}; namely, a (usually finite) set of quantum gates that can implement any general quantum computation.
Replacing gates by Hamiltonian terms, Ref.~\cite{UniversalHamiltonian} precisely defined analog Hamiltonian simulation, and what it means for a  family of Hamiltonians to be \emph{universal}. 
According to their definition, for any target local Hamiltonian $H$, there should exist a Hamiltonian $H'$ in the family that can simulate $H$. 
The ability to implement these universal families thus enables analog simulation of all local Hamiltonians. 
Ref.~\cite{UniversalHamiltonian} proves that various simple families of quantum spin-lattice models in two dimensions with tunable nearest-neighbor interaction energy are universal.
More families were shown to be universal in this sense by later works~\cite{Piddock2018Universal, Piddock2020,  Kohler2020}.
Importantly, it was shown in Ref.~\cite{UniversalHamiltonian} that analog simulations can reproduce all physical properties of the target system---including time-evolution, thermal states, and effects of local noise processes---to any desired precision.

\begin{table*}[htb]
    \centering
    \begin{tabular}{c|c|c|c|c}
         & ~spatial dimension~ & ~interaction energy~ & ~particle number~ & ~translation-invariance \\ \hline
         Cubitt et al.~\cite{UniversalHamiltonian} & 2D & exp(poly($n$))  & poly($n$) & semi \\
         Piddock-Bausch \cite{Piddock2020} & 2D & exp(poly($n$)) & exp(poly($n$)) & full \\
         Kohler et al. \cite{Kohler2020} & 1D & exp(poly($n$)) & poly($n$) & full* \\
         Kohler et al. \cite{Kohler2020} & 1D & exp(poly($n$)) & exp(poly($n$)) & full \\
         \hline
         This work (Theorem~\ref{thm:main}) & 2D & poly($n$) & poly($n$) & semi \\
         This work (Theorem~\ref{thm:1D-universal}) & 1D & poly($n$) & poly($n$) & none
    \end{tabular}
    \caption{The properties of currently known constructions of universal families of Hamiltonians that can simulate any local $n$-qudit Hamiltonian.
    Note the 1D construction by Kohler et al.\,\cite{Kohler2020}  with full* translation-invariance uses a Hamiltonian interaction that changes depending on the target Hamiltonian.
    }
    \label{tab:known}
\end{table*}

Nevertheless, a problem in the constructions given in all above-mentioned works~\cite{UniversalHamiltonian,Piddock2018Universal, Piddock2020,  Kohler2020} is that they are not efficient for general target Hamiltonians.
The efficiency in fact depends on the connectivity, or spatial dimensionality, of the target.
Any target Hamiltonian $H$ set on a 1D or 2D lattice can be simulated by a universal Hamiltonian simulator set on a 2D lattice, incurring only a polynomial overhead in both the number of particles and the interaction energy. However,  if the target $H$ is embedded in a higher dimension (e.g., when $H$ is 3D or has all-to-all interactions), an exponential overhead is required, either in the interaction energy or the number of particles simulating the interactions using gadgets~\cite{CaoNagajGadget}, which also gives up spatial locality. 

We call such families, which can reproduce the physics but not guarantee the efficiency in simulating general local Hamiltonians, \emph{weakly universal}. 
Recall that the whole point behind analog Hamiltonian simulations (which also originally motivated Feynman) is to avoid the exponential overhead that one incurs in classical simulations of many-body quantum systems.

In this work, we overcome the problem of exponential overhead and arrive at what we call {\it strongly universal} Hamiltonian simulators, which guarantee efficient simulation of all target local Hamiltonians, regardless of their geometry. 
We believe this is perhaps a more suitable definition of universality. 
Specifically, we provide a construction of analog simulators that requires only polynomial overhead in both the number of particles and the interaction energy, and enables simulation in 2D of any target local Hamiltonian regardless of its connectivity (Theorem~\ref{thm:main}).
In fact, the construction can be done using 2D spin-lattice models which include only a single type of nearest-neighbor interaction, 
where spatial variation appears solely in the interaction energies---a property we call \emph{semi-translation-invariance}.
As a consequence, the semi-translation-invariant 2D families of Hamiltonians found to be weakly universal in Ref.~\cite{UniversalHamiltonian} are in fact also strongly universal when more efficient constructions are utilized.
We further demonstrate that strong universality is achievable in 1D  (Theorem~\ref{thm:1D-universal}), albeit at the cost of spatially varying interaction types; while not semi-translation-invariant, this constructed simulator remains efficient.
The key technical advance enabling these results is a non-perturbative Hamiltonian-circuit-Hamiltonian mapping that efficiently simulates all-to-all interactions with spatially local couplings.
Table~\ref{tab:known} compares the resources required for simulating general Hamiltonians by our constructions to previous Hamiltonian simulator constructions.

Our work significantly expands the applicability domain of simple analog Hamiltonian simulators. In particular, we show that the aforementioned families of 1D and 2D universal analog simulators can efficiently simulate any Hamiltonian that has an efficient digital quantum simulation, i.e., any Hamiltonian whose time-evolution $e^{-iHt}$ has a polynomial-sized circuit approximation.
This includes fermionic Hamiltonians as well as Hamiltonians with $\poly(n)$ many $O(\log n)$-local terms which are digitally simulable
via Trotter decomposition and the Solovay-Kitaev algorithm~\cite{DawsonNielsen}.
Both classes are beyond the class of $O(1)$-local Hamiltonians targeted by the original definition of universal simulators in Ref.~\cite{UniversalHamiltonian}. 
More generally, since any physically realizable physical system can be simulated using a quantum circuit by the quantum Church-Turing thesis \cite{DeutschQCT,KayeLaflammeMoscaBook}, our results shows that our analog 1D and 2D simulators can efficiently simulate any physically realizable quantum system. 
Finally, we clarify the relationship between analog simulation and Hamiltonian dynamics simulation, which only simulates the evolution $e^{-iHt}$ applied to any input state. We show that while analog simulation implies dynamics simulation, the converse does not hold; for completeness, we also present universality results for dynamics simulation.

\section{Results}

\subsection{Theory of analog Hamiltonian simulation}

We start by describing the well-motivated definition of Ref.~\cite{UniversalHamiltonian} for analog simulation, 
which posits that ``$H'$ simulates $H$'' if the full spectrum of $H$ can be encoded as the low-lying part of the spectrum of $H'$, by an encoding that preserves locality of observables.
More precisely:
\begin{defn}[Local encoding, adapted from \cite{UniversalHamiltonian}]
\label{defn:local-encoding}
Consider an encoding map $\eps: \mathrm{Herm}((\CC^{d})^{\otimes n}) \to \mathrm{Herm}((\CC^{d'})^{\otimes n'})$ that takes Hermitian operators on $n$ qudits ($d$-dimensional particles) into operators acting on $n'$ $d'$-dimensional particles (where $d^n\le {d'}^{n'}$). 
We say $\eps$ is an \emph{$\eta$-local encoding} if we can write
\begin{equation} \label{eq:localenc}
\eps(A)=V(A\otimes P+\bar{A}\otimes Q)V^\dag,
\end{equation}
such that $V$ is an isometry satisfying $\|V-V_{\rm loc}\|\le \eta$ for some local isometry $V_{\rm loc}=\bigotimes_{i=1}^n V_i$, where each $V_i$ is an isometry acting on one qudit in the original system.
$P$ and $Q$ are locally orthogonal projectors, i.e., $P$ and $Q$ are orthogonal projectors and  $\forall i$ $\exists$  orthogonal projectors $P_i, Q_i$ acting on the same subsystem as $V_i$ such that $P_iQ_i=0$, $P_iP=P$ and $Q_iQ=Q$  (see also \cite[Supporting Information, Theorem 15]{UniversalHamiltonian}).
$\bar{A}$ is the complex conjugation of $A$. $\|\cdot\|$ is the spectral norm.

When $\eta=0$, we say $\eps$ is a \emph{local encoding}.

Furthermore, $\eps^\state$ is an \emph{$\eta$-local state-encoding} if it is of the form $\eps^\state(\rho)=\eps'(\rho)/\Tr(\eps'(\rho))$ for some $\eta$-local encoding $\eps'$.
For a given local encoding, one natural choice of the corresponding local state-encoding would set $\eps'=\eps$. An alternative choice of local state-encoding could take $V'=V$, $P'=\sigma$, and $Q'=0$ where $\sigma$ is a state satisfying $P\sigma=\sigma$.
\end{defn}
\begin{defn}
[Analog Hamiltonian simulation, adapted from \cite{UniversalHamiltonian}]
\label{defn:CMPsimul}
A Hamiltonian $H'$ is a $(\Delta, \eta,\epsilon)$-\emph{simulation} of an $n$-qudit Hamiltonian $H$ if there exists an $\eta$-local encoding $\eps_\eta$ as in Definition~\ref{defn:local-encoding}  such that
\begin{enumerate}
\item $\eps_\eta(\Id) = P_{\le \Delta(H')}$; 
\item $\|H'_{\le \Delta} - \eps_\eta(H)\| \le \epsilon$.
\end{enumerate}
Here, $P_{\le \Delta(H')}$ is the projector onto the subspace of eigenstates of $H'$ with eigenvalue $\le \Delta$, and
$H'_{\le \Delta} =  P_{\le \Delta(H')} H'$ is the restriction of $H'$ onto these states. 
We say the simulation is \emph{efficient} if both the number of particles in $H'$ and its maximum energy $\|H'\|$ are at most $O(\poly(n,\eta^{-1},\epsilon^{-1}, \Delta))$,  
and the description of $H'$ is efficiently computable.
\end{defn}
Under this definition, Ref.~\cite{UniversalHamiltonian} showed that implementing $H'$ allows one to approximately reproduce all physical properties of $H$,
implying that the term ``simulation'' means essentially reproducing any physical aspect. 
Specifically, since $\eps_\eta$ is an $\eta$-local encoding, all local observables $A$ with respect to $H$ are mapped to local observables $\eps_0(A)$ for $H'$ up to error $\eta$. 
Correspondingly, there is a local state-encoding $\eps^{\rm state}_0(\rho)=\eps_0(\rho)/\Tr[\eps_0(\rho)]$ that satisfies $\Tr(A\rho) = \Tr[\eps_0(A)\eps^{\rm state}_0(\rho)]$. 
Gibbs states (thermal ensembles) of $H$ are mapped to Gibbs states of $H'$, with errors exponentially suppressed by the energy cutoff $\Delta$. 
Time-evolution $e^{-iHt}$ applied to any initial state $\rho$ can also be simulated by $e^{-iH't}$ applied on the appropriately encoded state $\eps^{\rm state}_0(\rho)$, and the error in this simulation grows as $O(\epsilon t + \eta)$.

Additionally, it has been argued that such analog simulators may not require active error-correction~\cite{CiracZollerNatPhys2012}.
This is made more explicit in Ref.~\cite{UniversalHamiltonian}, which argued that this is indeed true under Definition~\ref{defn:CMPsimul} and a reasonable assumption that the noise causes no significant leakage into the high energy subspace separated by a sufficiently large $\Delta$.
Then any local noise channel $\cN'$ acting on the simulator system can be approximated as an encoded version of a local noise channel $\cN$ in the original system: $\|\cN'(\eps^{\rm state}_0(\rho)) - \eps^{\rm state}_0(\cN(\rho))\|_1 \le O(\eta)$ (see \cite[Corollary 33 in SI]{UniversalHamiltonian}).
Even when the encoded noise process is not the same as the true one, the noisy simulation by $H'$ may still allow one to probe, without error-correction, the relevant physics of $H$ that are insensitive to details of the noise processes.

Our goal is to understand which families of Hamiltonians can be used to simulate all other physical Hamiltonians as in Definition~\ref{defn:CMPsimul}, with only polynomial overhead in all resources, for all target local Hamiltonians. 
To this end we define: 
\begin{defn}[Weak and strong universality]
\label{defn:universality}
A family of Hamiltonians $\F = \{H_{m}'\}$ is \emph{weakly universal} if given any $\Delta, \eta,\epsilon>0$, any $O(1)$-local $n$-particle Hamiltonian $H$ can be $(\Delta,\eta,\epsilon)$-simulated by some $H_m'\in \F$.
Such a family is \emph{strongly universal}
if for all $H$ with $\|H\|\le\poly(n)$, the simulation is always efficient ---i.e., $H_m'$ is efficiently computable in $O(\poly(n,\eta^{-1},\epsilon^{-1}, \Delta))$ time from the description of $H$, requires $O(\poly(n, \eta^{-1},\epsilon^{-1}, \Delta))$ particles, and $\|H_{m}'\| = O(\poly(n,\eta^{-1},\epsilon^{-1}, \Delta))$.
\end{defn}

We remark that this notion of strong universality is equivalent to the ``efficient universality'' used in prior works (e.g., \cite{UniversalHamiltonian}). We adopt the ``strong/weak'' terminology here to emphasize that some universal families lack a known polynomial bound on resources. Note that weak universality does not mean that the simulations are provably inefficient.

Following Ref.~\cite{UniversalHamiltonian}, we consider Hamiltonian simulators that only involve up to 2-local interactions drawn from a set $\cS=\{h_\alpha\}_\alpha$.
Note this set $\cS$ may contain only a single term.
Then we define an $\cS$-Hamiltonian simulator to be a Hamiltonian of the following form:
\begin{equation}\label{eq:simulationform} 
H' = \sum_{\braket{i,j}\in E} J_{ij} h_{\alpha_{ij}}^{(i,j)}, \quad
\text{where}
~
\|h_{\alpha_{ij}}\|\le 1 \text{ and } J_{ij}\in \mathbb{R}.
\end{equation}
Here, $E$ is some set of edges describing the connectivity of the qudits,
$h_{\alpha_{ij}}^{(i.j)}$ is some two-body operator $h_{\alpha_{ij}} \in \cS$ acting on qudit $i$ and $j$, and $J_{ij}$ is the interaction energy.%

It is shown in Ref.~\cite{UniversalHamiltonian} that many families of $\cS$-Hamiltonians on qubits are weakly universal, even when restricting the connectivity $E$ of the Hamiltonian simulator to the 2D square lattice. 
Two examples are the Heisenberg interaction ($\cS = \{X\otimes X+Y\otimes Y+Z\otimes Z\}$) and the XY-interaction ($\cS =  \{X\otimes X + Y\otimes Y\}$).
Here and below, we denote $(X,Y,Z)=(\sigma_x,\sigma_y,\sigma_z)$ as the Pauli matrices.
With these choices of $\cS$, the Hermitian terms in Eq.~\eqref{eq:simulationform} are all equal up to their relative weights.
To state this more concisely, we define:
\begin{defn}[Full and Semi-Translation-Invariance]
A Hamiltonian $H'$ has \emph{semi-translation-invariance (or semi-TI)} if all two-body operators are the same up to the scaling by $J_{ij}$, i.e., $h_{\alpha_{ij}}^{(i,j)}= h^{(i,j)}$.
We say $H'$ has \emph{full translation-invariance  (or full-TI)} if it has semi-translation-invariance and all the interaction energies are the same, i.e., $J_{ij} = J$.
\end{defn}

Ref.~\cite{UniversalHamiltonian} shows that any family of $\cS$-Hamiltonians is weakly universal as long as $\cS$ is non-2SLD, which means that the 2-local parts of all the interactions in $\cS$ are not simultaneously and locally diagonalizable: 
\begin{defn}[2SLD interactions~\cite{UniversalHamiltonian}]
\label{def:2SLD}
Suppose $\cS$ is a set of interactions on 2 qubits.
We say $\cS$ is \emph{2SLD} if there exists $U\in\mathrm{SU}(2)$ such that for each $H_i\in \cS$, $U^{\otimes 2} H_i (U^\dag)^{\otimes 2} = \alpha_i Z\otimes Z + A_i\otimes \Id + \Id\otimes B_i$, where $\alpha_i\in \mathbb{R}$ and $A_i,B_i$ are 1-local operators.
Otherwise, $\cS$ is \emph{non-2SLD}.
\end{defn}

The main result of Ref.~\cite{UniversalHamiltonian} is then:
\begin{thmCMP}[\cite{UniversalHamiltonian}]
\label{thm:CMP}
Any $\cS$-Hamiltonian family set on a $2D$ square lattice of qubits is weakly universal as long as $\cS$ is non-2SLD.
\end{thmCMP}

There are known non-2SLD sets of interactions $\cS$ which contain a single interaction term (e.g., Heisenberg or XY interactions), thus semi-translation-invariant families of Hamiltonians in 2D that are weakly universal exist. 
Later works have extended this result to show weak universality of various families that use qudits~\cite{Piddock2018Universal}, are embedded in higher dimensions~\cite{Piddock2020}, or are fully translation-invariant in 2D or 1D~\cite{Piddock2020,Kohler2020}.

A major question remains: Can the simulations be made efficient, so that universality is achieved in the strong sense? 
As mentioned earlier, the constructions used by Ref.~\cite{UniversalHamiltonian} to prove Theorem \hyperref[thm:CMP]{CMP18} are only efficient if the target Hamiltonian have the same or lower spatial dimensionality as the simulator Hamiltonian.
When attempting to simulate a 3D target Hamiltonian (or worse, all-to-all interacting, such as the SYK model \cite{SYK1,*SYK2}) by a 2D simulator, however, the simulation requires exponentially large interaction energy $J_{ij} = 2^{\Omega(\poly(n))}$.
Similar exponential overhead is required in Refs.\,\cite{Piddock2018Universal, Piddock2020, Kohler2020} for simulating local Hamiltonians with general connectivity.

\subsection{Strongly universal Hamiltonian simulators}

With the definitions and background introduced, we now show how to use the families of 2D Hamiltonians proposed in Ref.~\cite{UniversalHamiltonian} to achieve not only \emph{weak universality} but also \emph{strong universality} for analog simulation.

\begin{thm}
\label{thm:main}
Any $\cS$-Hamiltonian family on the 2D square lattice is strongly universal, as long as $\cS$ is non-2SLD.
In particular, it's sufficient for $\cS$ to contain only a single interaction (such as Heisenberg or XY-interaction), implying that there are semi-translation-invariant Hamiltonians in 2D that are strongly universal. 
\end{thm}

We further show that strong universality can even be achieved by 1D Hamiltonians with nearest-neighbor interactions. 
However, we give up translation-invariance in the process.
\begin{thm} \label{thm:1D-universal}
There is a strongly universal family of 1D Hamiltonians consisting of nearest-neighbor interactions acting on a line of 8-dimensional particles.
\end{thm}

The interactions used in Theorem \ref{thm:1D-universal} arise from the circuit-to-Hamiltonian construction applied to a quantum circuit in 1D.
These interactions are tailored to represent the gates that make up a general 1D circuit, and thus they do not possess any translation-invariance.

To prove our results, we note that previous constructions \cite{UniversalHamiltonian,Piddock2018Universal,Piddock2020,Kohler2020} incurred an exponential overhead because they relied purely on perturbative gadgets to reduce the connectivity of qudits.
Specifically, in these constructions, in order to reduce degree in the interaction graph from $\Omega(n)$ to $O(1)$ so that the Hamiltonian can be embedded on a finite-dimensional lattice, it is necessary to apply $\Omega(\log n)$ rounds of perturbative gadgets, each of which roughly halves the degree.
Since the required interaction energy increases polynomially for each application of the perturbation gadget (i.e., $J \longmapsto [J \poly(n)]^c$ for some constant $c > 1$), the final Hamiltonian requires interaction energy scaling as $J_{\rm final} = n^{c^{\Omega(\log n)}} = 2^{\Omega{(\poly(n))}}$ in these constructions. 

To circumvent this problem, we start from an idea from Ref.~\cite{sparsify} (which was recently applied in Ref.~\cite{Kohler2020})
in which a transformation is performed by first mapping the target Hamiltonian $H$ to a quantum circuit (implementing a phase estimation algorithm with respect to $e^{-iHt}$) and only then mapping back to another Hamiltonian $H'$ using a variant of the circuit-to-Hamiltonian construction~\cite{KSV02}.
The reason to use the intermediate quantum circuit step is that unlike Hamiltonians, circuits can be straightforwardly made ``sparse"---i.e., each qubit in the circuit is only acted upon by a few gates.
This can be done by swapping qubits to fresh ancilla qubits after every computational gate.
The sparsity of the circuit is what enables the final Hamiltonian to be embeddable in a 1D or 2D lattice.

Our construction of Hamiltonian simulators in 2D first utilizes techniques from earlier Hamiltonian complexity literature~\cite{AharonovAQCUniversal,OliveiraTerhal}. To obtain a 
semi-translation-invariant simulators, we borrow additional gadgets from Ref.~\cite{UniversalHamiltonian}.
To obtain a 1D Hamiltonian simulator, we employ a construction of QMA-complete 1D Hamiltonians from~\cite{AharonovLine2009,Hallgren} to simulate the quantum circuit using nearest-neighbor Hamiltonian interactions on a line of 8-dimensional particles. 
Although na\"ive application of these methods creates issues such as nonlocality of the encoding in the simulation, we are able to address them by carefully modifying the constructions.
See the Methods section for more details on the proofs.

While our discussion thus far is restricted to simulating target Hamiltonians that are $O(1)$-local, the universal families we have constructed can actually simulate more complicated Hamiltonians.

\begin{defn}[digitally simulable] \label{defn:circuit-sim}
An $n$-qudit Hamiltonian $H$ is \emph{digitally simulable} 
if for any $\epsilon >0$ and $t$ one can construct a quantum circuit $U$ consisting of $\poly(n,t, \epsilon^{-1})$ one- or two-qubit gates (drawn from some universal gate set) such that $\|U - e^{-i Ht} \| \le \epsilon$, and the description of $U$ 
is computable in $\poly(n, t, \epsilon^{-1})$ time.
\end{defn}

Since the aforementioned efficient simulators are based on circuit-to-Hamiltonian constructions, they easily generalize to any digitally simulable Hamiltonian:
\begin{thm}
\label{coro:equiv}
Any digitally simulable Hamiltonian has an efficient analog simulation by a Hamiltonian from any of the strongly universal families described in Theorems~\ref{thm:main} and \ref{thm:1D-universal}.
\end{thm}

\subsection{Analog vs. dynamics simulation}
\label{sec:dynamics}

The results above refer to analog Hamiltonian simulation as defined in Definition~\ref{defn:CMPsimul}. Nevertheless, there is an alternative, commonly adopted notion of Hamiltonian simulation which focuses only on simulating time-evolution dynamics.
While at first the distinction may appear negligible, there are in fact important differences between the two notions.
We define dynamics simulation, borrowing inspiration from Definition~\ref{defn:CMPsimul}, as:
\begin{defn}[dynamics simulation]
\label{defn:dynamics}
A Hamiltonian $H'$ is an \emph{$(\eta, \epsilon)$-dynamics simulation} of an $n$-qudit Hamiltonian $H$ if for any time $t\ge 0$, there exist (possibly time-dependent) encoding maps $\eps_{\rm in}^\state(\rho,t)$ and $\eps_{\rm out}^\state(\rho, t)$ that are $\eta$-local state-encodings as in Definition~\ref{defn:local-encoding}, such that for any density matrix $\rho_0$, we have
$\|e^{-iH't'}\eps_{\rm in}^\state(\rho_0, t) e^{iH' t'} - \eps_{\rm out}^\state(e^{-iHt}\rho_0 e^{iHt}, t)\|_1 \le \epsilon t$, where $t' = \poly(t,n)$.
\end{defn}

\begin{figure}[tb]
    \centering    \includegraphics[width=0.9\linewidth]{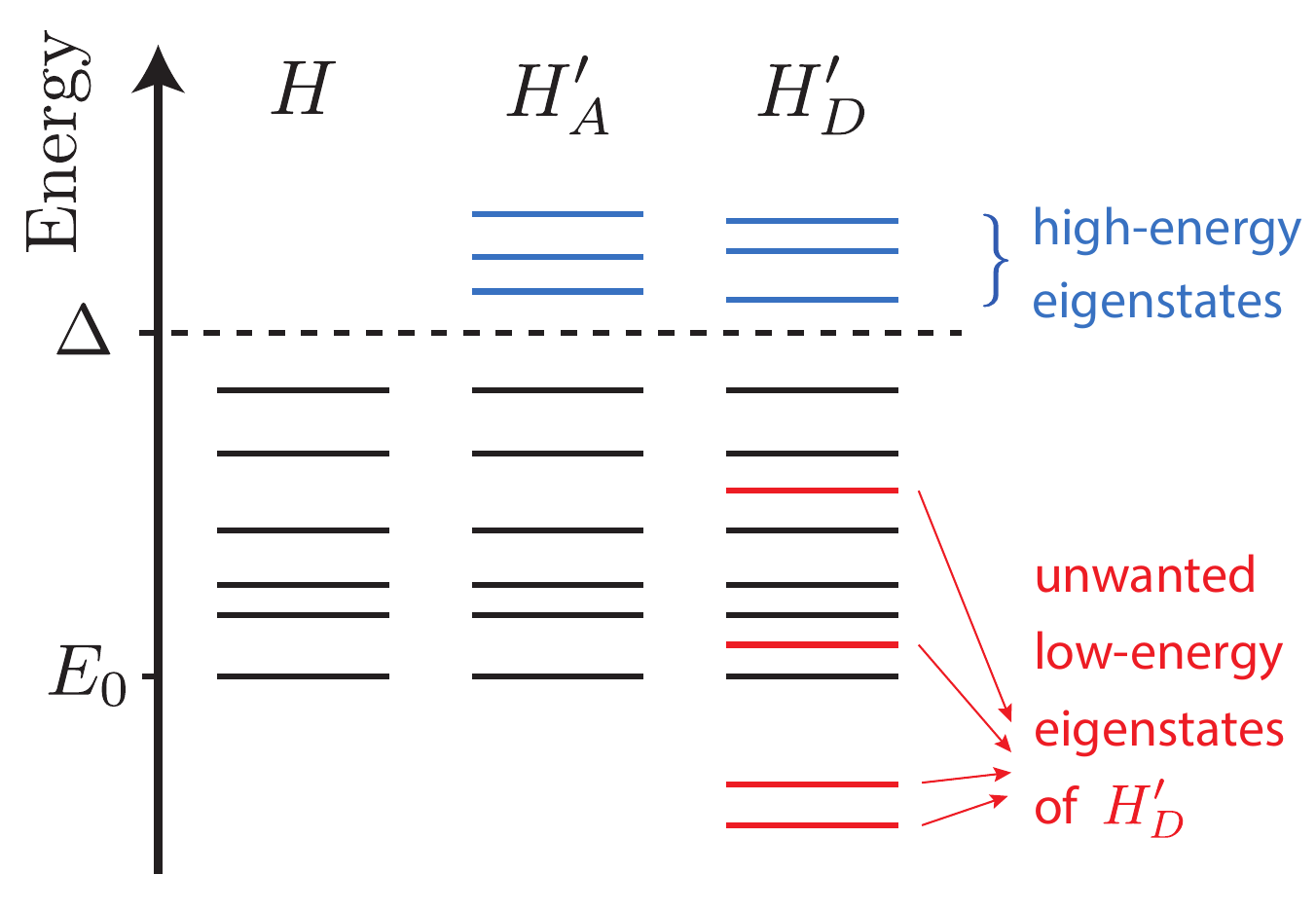}
    \caption{\textbf{Sketch of the spectra of a Hamiltonian $H$, its analog simulator $H'_{A}$, and its dynamics simulator $H'_{D}$.} The black eigenstates of $H'_{A}$ and $H'_{D}$ are those in the range of $\eps^\state(\Id)$ and encode the eigenstates of $H$. In contrast to the analog simulator $H'_{A}$, however, the dynamics simulator $H'_{D}$ may also contain other eigenstates, such as the unwanted low-energy eigenstates (red). Leakage into these states, possibly driven by thermal fluctuations, can cause $H'_D$ to fail to be a good analog simulator. Nevertheless, if the leakage can be suppressed, then $H'_D$ restricted to the range of $\eps^\state(\Id)$ (black) and the high-energy eigenstates (blue) may still serve as an analog simulation of $H$.}
    \label{fig:dyn-simulator}
\end{figure}
See Ref.~\cite{BB17} for a related definition, which allows $\eps^{\rm in}_\state$ and $\eps^{\rm out}_\state$ to be any polynomial-sized quantum circuit.

As we shall see, dynamics simulation is a somewhat weaker notion than analog simulation.
We first show that an analog simulator is automatically a dynamics simulator; a somewhat restricted version of the converse statement also holds:
\begin{claim}
\label{claim:dyn-vs-analog}
If $H'$ is a $(\Delta,\eta,\epsilon)$-analog simulation of $H$ as in Definition~\ref{defn:CMPsimul}, then $H'$ is an $(\eta,2\epsilon)$-dynamics simulation of $H$.
Conversely, if $H'$ is an $(\eta, 0)$-dynamics simulation of $H$ with time-independent encodings $\eps_{\rm in}^\state(\rho,t)=\eps_{\rm out}^\state(\rho,t)=\eps^\state(\rho)$ and $t'=t$, then $H'+c I$ restricted to a certain subspace is a $(\Delta, \eta, 0)$-analog simulation of $H$, for some offset $c$ and some $\Delta \ge \|H\|$.
\end{claim}
See \cref{supp1} for a detailed proof.

A dynamics simulator need not function as an analog simulator of the full Hamiltonian because it can possess unwanted low-energy eigenstates without counterparts in the target system. An example is illustrated as $H_D'$ in Fig.~\ref{fig:dyn-simulator}.
Such a dynamics simulator only becomes an analog simulator after restriction to the desired eigenstates; however, any uncontrolled leakage into the unwanted eigenstates could cause the analog simulation to fail.
In particular, thermal fluctuations and other noise sources can drive the dynamics simulator to populate these low-energy eigenstates, preventing it from faithfully reproducing the equilibrium thermodynamical properties of the target Hamiltonian.
Moreover, a dynamics simulator may allow a nonlinear time parametrization $t'=f(t)$ that can distort the spectrum significantly and further obstruct faithful analog simulation.

Thus, dynamics simulation is strictly less powerful than analog simulation. 
It is interesting to explore the notion of universality in the context of dynamics simulation.
We define weak and strong universality for dynamics simulator by slightly modifying Definition~\ref{defn:universality} in the obvious way. 
Then Claim~\ref{claim:dyn-vs-analog} implies that the families of Hamiltonians in Theorems \ref{thm:main} and \ref{thm:1D-universal}, which are strongly universal for analog simulations, are also strongly universal for dynamics simulation.

\begin{coro}
Any $\cS$-Hamiltonian family on the 2D square lattice is strongly universal for dynamics simulation of any Hamiltonian, as long as $\cS$ is non-2SLD.  When $\cS$ contains only a single interaction, we have semi-translation-invariant families of Hamiltonians in 2D that are strongly universal for dynamics simulation.
There is also a family of 1D Hamiltonians for 8-dimensional particles that is strongly universal for dynamics simulation (but with no semi-translation-invariance).
\end{coro}

We remark that strong universality for dynamics simulation is easier to achieve than for analog simulation.
For example, an alternative proof of strong universality of 2D dynamics simulator, can also be shown from previous results regarding universal quantum computation using quantum walk Hamiltonians\,\cite{MPQW}.  
Furthermore, Ref.~\cite{BB17} has constructed a family of 1D, fully translation-invariant Hamiltonians based on quantum cellular automata on 14580-dimensional particles that is strongly universal for dynamics simulation, albeit with a non-local encoding.
Nevertheless, these Hamiltonians are not analog simulators due to the presence of exponentially many low-energy states corresponding to other computations that can be encoded in the initial state of the automaton.
Hence, such simulators likely exhibit less noise-resilience.
We do not know how to generalize Ref.~\cite{BB17}'s construction to achieve strongly universal analog simulators with full translation-invariance.

\section{Discussion}

In this work, we have significantly improved the prospects of universal analog quantum simulation by showing that efficient universal simulation is possible with simple families of Hamiltonians embedded in low-dimensional space.
Unlike the exponential overheads generally required in previous works~\cite{UniversalHamiltonian,Piddock2018Universal,Piddock2020,Kohler2020}, our results show that only polynomial overheads in both particle number and interaction energy are sufficient to simulate any digitally simulable Hamiltonian with arbitrary connectivity by some universal Hamiltonians embedded in 1D or 2D.
We note that the polynomial overhead in the interaction energy cannot be brought down to a constant using finite-dimensional Hamiltonian simulators, by an earlier impossibility result~\cite{sparsify}. 
We remark that even though there are weakly universal, fully translation-invariant families of Hamiltonians in 1D and 2D~\cite{Piddock2020,Kohler2020}, the interaction energy in those Hamiltonians scales exponentially with the size of the target Hamiltonian.

We have also clarified the difference between analog Hamiltonian simulation and dynamics simulation.
Although often referred to interchangeably, we show that analog simulation is a strictly more powerful notion than dynamics simulation, which only aims to reproduce the time-evolution under the Hamiltonian.
Strongly universal dynamics simulators are also easier to construct than their analog counterparts.
For example, we can show that 1D Hamiltonians with full translation-invariance are strongly universal for dynamics simulation, but not yet for analog simulation.

An independent concurrent work \cite{Kohler2022general} proved that a family of Hamiltonians is strongly (or in their terminology, efficiently)  universal if and only if it is closed under addition and QMA-complete under what they call faithful reductions. They showed that the circuit-to-Hamiltonian construction used in Ref.~\cite{KSV02} to prove QMA-completeness of general local Hamiltonians (without restriction on geometry) can be modified to satisfy these conditions. Their proof strategy for showing faithfulness bears some similarities to our approach (e.g., idling trick).
To check that their condition holds for the Hamiltonian families considered in the current paper, one would need a new proof that these 2D and 1D Hamiltonians are QMA-complete under faithful reduction; we believe that this can be done by adopting arguments in our proof of Proposition~\ref{prop:sparseHam} and Theorem~\ref{thm:1D-universal}. This would provide a somewhat different proof of the strong universality of the Hamiltonian families in Theorems~\ref{thm:main} and \ref{thm:1D-universal}.

Several natural open questions remains for future work. Are there strongly universal semi-translation-invariant analog Hamiltonian simulators in 1D?
The known gadgets to simulate general interactions with a single type of interaction seem to require ancilla particles placed in more than 1D; hence, new gadgetry appears necessary.
Are there strongly universal Hamiltonians with full translation-invariance in any constant dimensions? Such constructions would be very helpful for near-term implementations.
We note this is impossible when the only free parameter of the Hamiltonians is the number of particle $n'$; since such Hamiltonians can be described by $O(\log n')$ bits of information, they cannot represent general Hamiltonians on $n$ qudits that are described by $\poly(n)$ bits unless $n'=\exp(\poly (n))$.
Indeed, Refs.~\cite{Piddock2020,Kohler2020} constructed such families of Hamiltonians that are weakly universal, but cannot be strongly universal;
this also implies that not all weakly universal families are strongly universal.

To obtain full-TI strongly universal analog simulators, we may consider relaxing translation-invariance by allowing more free parameters in the interaction operator. This allows us to encode general target Hamiltonians using the $\poly(n)$ bits describing the interaction.
Ref.~\cite{Kohler2020} has done this to maintain $O(\poly(n))$ particles overhead in the simulator, but their construction still requires exponential overhead in the interaction energy.
Nevertheless, it is likely possible to efficiently simulate any full-TI Hamiltonian by a family of full-TI Hamiltonians, where the description of both Hamiltonians are $O(\log n)$ bits.
It would be worthwhile to investigate whether these constructions can be improved, or otherwise prove impossibility of strong universality of full-TI Hamiltonians.

While this work constructs efficient universal analog quantum simulators 
using simple 1D or 2D systems, with only polynomial overheads in all resources, the constructions presented here are not optimized for actual implementation, and there is much room for improvement for practical applications.
The main contributor to the resource overhead in our analysis is the non-perturbative Hamiltonian-circuit-Hamiltonian pipeline, which is currently dominated by the Trotter decomposition of $e^{-iH t}$ in the phase-estimation subroutine and amplified by the energy penalties involved in the circuit-Hamiltonian construction.
These overheads could potentially be improved by replacing Trotterization with QSVT-based~\cite{QSVT2019} method which has an exponentially better dependence on the precision parameter.
Furthermore, we can consider a hybrid-approach that applies the costly non-pertubative mapping to only the high-degree or high-locality terms in the input Hamiltonian while handling the rest with cheaper perturbative gadgets.
On the hardware side, superconducting qubit platforms~\cite{Andersen2025thermalization} with native or programmable Heisenberg/XY couplings are well-suited to implement our 2D universal Hamiltonians; it would also be interesting to construct other universal families of Hamiltonians tailored to neutral atom and trapped ion platforms~\cite{Browaeys2020atoms,Monroe2021ions} whose geometries are more flexible.
More generally, there are useful trade-offs between structural constraints and resource costs; for example, relaxing semi-translation-invariance can reduce resource overheads by dispensing with gadgets that enforce a single interaction type.
We further anticipate that leveraging other methods, such as the recently proposed dissipative gadgets~\cite{Harley2024going}, may achieve simulation with better resource scaling.
As experimental realizations of analog quantum simulators develop rapidly, we hope our work helps to expand their scope to simulate all physical systems and tackle classically intractable problems.

\section{Methods}

Here we sketch the construction of universal Hamiltonian simulators for proving Theorem~\ref{thm:main} and Theorem~\ref{thm:1D-universal}.

\subsection{Overview}

The starting point of our proofs for both Theorem~\ref{thm:main} and \ref{thm:1D-universal} is as follows. 
We note that previous constructions \cite{UniversalHamiltonian, Piddock2018Universal, Piddock2020, OliveiraTerhal} relied purely on perturbative gadgets to reduce the connectivity of qudits. 
As mentioned earlier in the Results section, reducing the degree in the interaction graph from $\Omega(n)$ to $O(1)$ required $\Omega(\log n)$ rounds of perturbative gadget, yielding a final Hamiltonian with interaction energy blowing up as $2^{\Omega(\poly(n))}$.

Instead, we reduce the connectivity of the qudits in our construction by first mapping the target Hamiltonian $H$ to a quantum circuit performing the phase estimation algorithm with respect to $e^{-iHt}$. 
Using standard techniques including Trotter decomposition~\cite{Trotter}, we can embed this circuit in 1D, using only nearest-neighbor gates on a line of qubits, while still applying the desired phase estimation with sufficient accuracy.
For any given energy eigenstate $\ket{\psi_E}$, where $H\ket{\psi_E}=E\ket{\psi_E}$, this circuit writes down the energy eigenvalue as a bit-string in some ancilla qubits to approximately yield  $\ket{\psi_E}\otimes\ket{E}$.
This is the first step, summarized in Proposition \ref{prop:PEcircuit}.

After the first step (Proposition \ref{prop:PEcircuit}), 
in which the target Hamiltonian is replaced by a phase estimation circuit operating in 1D, we wish to map the resulting circuit back to a Hamiltonian that can be embedded in a 1D line or 2D lattice.

In the 2D case, we do this by applying ideas from Refs.~\cite{AharonovAQCUniversal, OliveiraTerhal} to convert the 1D circuit given in Proposition~\ref{prop:PEcircuit} to a 2D spatially sparse Hamiltonian.
Roughly speaking, spatial sparsity means that each qubit participates in only a constant number of local interactions or gates, arranged in a spatially local way on the 2D plane (see Definition~\ref{defn:spatial-sparsity}). 
Note that a circuit that uses only nearest-neighbor gates in 1D is not necessarily spatially sparse since each qubit can participate in $O(\poly(n))$ gates. 
Nevertheless, we can utilize swap gates to transform the 1D circuit into a 2D spatially sparse circuit non-perturbatively, where the second spatial dimension plays the role of time in the circuit.
Having enforced the 2D spatial sparsity of the circuit, we next apply the Feynman-Kitaev circuit-to-Hamiltonian construction \cite{KSV02} to obtain a 2D spatially sparse Hamiltonian $H_\circuit$ with roughly the same connectivity as the circuit. 
Importantly, this Hamiltonian has an exponentially large degeneracy of ground states: for each eigenstate of $H$, we have a groundstate of $H_\circuit$ corresponding to the computational history of the phase estimation circuit running on that eigenstate as input.

In order to restore the spectral features of $H$ in $H_\circuit$, we apply the additional trick of imposing bit-wise energy penalties on the energy bit-string ancilla qubits to match the energy of the eigenstate.
The states corresponding to the different computational histories of different eigenstates of $H$ exhibit unwanted entanglement between the original qubits and the ancilla qubits, which causes decoherence in the simulation.
For example, the state $\ket{\psi} = \sum_E \ket{\psi_E}$ in the original system is roughly mapped to $\ket{\tilde\psi} \approx \sum_E \ket{\psi_E}\otimes \ket{E}$.
This decoherence can be repaired by the tricks of ``uncomputing'' the unwanted entanglement with the ancilla, and ``idling'' at the end of the (virtual) circuit, so that $\ket\psi$ is instead mapped to $\ket{\psi'} \approx (\sum_E \ket{\psi_E})\otimes \ket{a}$ where $\ket{a}$ is some explicit state on the ancilla qubits.

This part of the construction, obtaining $H_\circuit$, is summarized in Proposition~\ref{prop:sparseHam}.

\begin{figure}[tb]
    \centering
    \includegraphics[width=\linewidth]{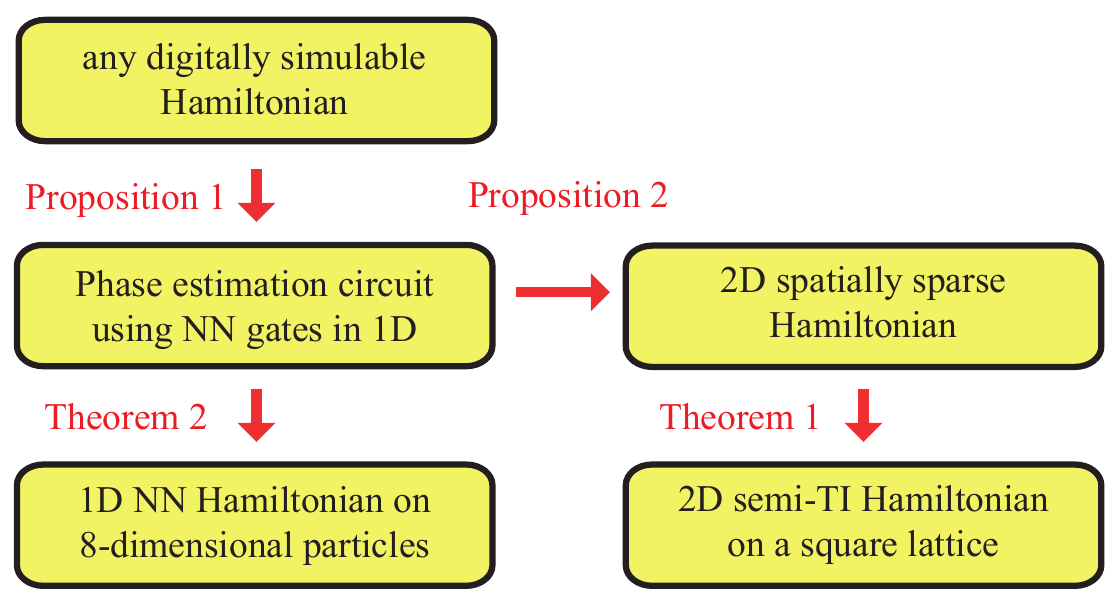}
    \caption{\textbf{Overview of our constructions of 1D and 2D universal families of Hamiltonians.} NN = nearest-neighbor. TI = translation-invariant.
    }
    \label{fig:UHconstruction}
\end{figure}

We then prove strong universality of 2D Hamiltonians by converting $H_\circuit$, which is spatially sparse in 2D, to a semi-translation-invariant Hamiltonian $H'$ on a 2D square lattice using gadgets from Ref.~\cite{UniversalHamiltonian}.
It can be shown that all steps of the construction incur only polynomial overhead in the interaction energy and the number of qubits, even if we restrict to 2D lattice Hamiltonians with Heisenberg or XY-type interactions.
Hence, such families of semi-TI 2D Hamiltonian are strongly universal, concluding our proof of Theorem~\ref{thm:main}.

When attempting to obtain a universal family of Hamiltonians in 1D, we cannot directly apply the above methods to the 1D circuit to obtain a 1D Hamiltonian.
Specifically, there are two difficulties: (1) embedding both the space and time dimension of the 1D circuit in 1D, and (2) enforcing semi-translation-invariance in 1D.
The proofs of the 1D and 2D cases thus diverge after the first step
of Proposition \ref{prop:PEcircuit}. 
Figure~\ref{fig:UHconstruction}  explains the proof structure for both 2D and 1D case. 

Instead, we construct a family of 1D Hamiltonians that simulate 1D phase estimation circuits by combining the tools of uncomputing and idling with previous circuit-to-Hamiltonian constructions that yielded QMA-complete 1D Hamiltonians~\cite{AharonovLine2009, Hallgren} to achieve our Theorem~\ref{thm:1D-universal}.
The interactions in this Hamiltonian family are nearest-neighbor operators that enforce a set of transition rules, so that the zero-energy eigenstates are those corresponding to performing the circuit computation correctly.
This address the first difficulty.
However, we are currently unable to enforce semi-translation-invariance, so the second difficulty is left open.
This is because the interactions here vary according to the gate sequence in the 1D circuit, and the gadgets from Ref.~\cite{UniversalHamiltonian} to obtain semi-TI only work in 2D. 
After applying the 1D circuit-to-Hamiltonian construction from Refs.~\cite{AharonovLine2009, Hallgren}, we recover the spectral properties of $H$ by imposing bit-wise energy penalties and repair decoherence via uncomputing and idling, similar to the 2D case.
We note a na\"ive application of their construction produces an encoding that is not local in the sense of Definition~\ref{defn:local-encoding}.
Nevertheless, we are able to restore the locality of encoding via a simple modification to the transition rules.

In both constructions of the 1D and 2D families, our non-perturbative techniques avoid the exponential blow-up in the required interaction energy, which brings the scaling of the energy overhead $J_\text{final}$ down to only $O(\poly(n))$.

Theorem~\ref{coro:equiv} follows since both our 2D and 1D constructions are efficient for  any input Hamiltonian $H$ as long as the phase estimation circuit on $e^{iHt}$ resulting from Proposition~\ref{prop:PEcircuit} is polynomial in size.
If $H$ is digitally simulable, then this is true by Definition~\ref{defn:circuit-sim}.

In the remainder of the Methods section, we provide more detailed sketches of the steps in our constructions.

\subsection{The 1D phase estimation circuit (Proposition \ref{prop:PEcircuit})} 

We first show---and this is fairly standard---that one can construct a phase estimation circuit $U_{\rm PE}^\NN$ that would take eigenstates of $H$ as input and (approximately) write down their energy on ancilla qubits, using only nearest-neighbor gates acting on a line of qubits.

Given any $n$-qubit digitally simulable Hamiltonian $H$, let us write the decomposition in its energy eigenbasis as $H=\sum_\mu E_\mu \ketbra{\psi_\mu}$, where $0\le E_\mu\le E_{\max}$ without loss of generality.
Note the upper bound $E_{\max}=O(\poly(n))$ can be computed without knowledge of the energy eigenvalues, e.g., by adding up the spectral norm of individual local terms of $H$.

Ideally, we want a phase estimation circuit $U_{\rm PE}^\ideal$ that acts on any input state of the form $\ket{\psi}\ket{0^m}=\sum_\mu c_\mu \ket{\psi_\mu}\ket{0^m}$ in the following way
\begin{align}
\label{eq:PE-ideal-state}
U_{\rm PE}^\ideal \sum_\mu c_\mu\ket{\psi_\mu}\ket{0^m} =  \sum_\mu c_\mu\ket{\psi_\mu}\ket{E_\mu},
\end{align}
where $\ket{E_\mu}=\ket{\varphi_{\mu,1}\varphi_{\mu,2}\dots}$, and $\varphi_\mu=0.\varphi_{\mu,1}\varphi_{\mu,2}\dots$ is the binary representation of the real number $\varphi_\mu = E_\mu/E_{\max}$.

In reality, however, there are various approximations we must make to construct the phase estimation circuit in 1D.
First, we cannot get infinite precision in the estimation of $E_\mu$, nor can we get an efficient approximation to within exponential precision (polynomially many bits).
Hence, we restrict the first $s=O(\log(n))$ qubits of the $m$-qubit ancilla register to encode the energy $E_\mu$ as $\ket{\tilde{E}_\mu}=\ket{\varphi_1\varphi_2\dots\varphi_s}\otimes\ket{\text{rest}_\mu}$.
Here $\ket{\rest_\mu}$ is some arbitrary state in the remaining $m-s$ ancilla qubits.

\begin{figure}[t]
\includegraphics[width=\linewidth]{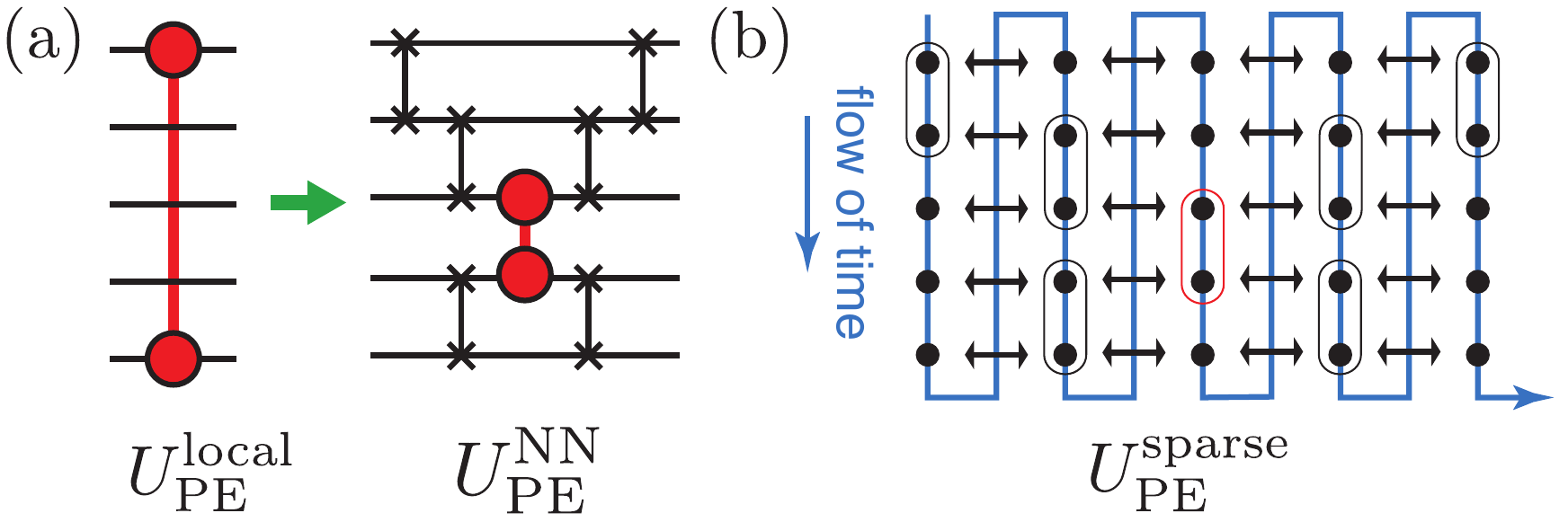}
\caption{\textbf{Illustration of converting any circuit to a spatially sparse one.} (a) Adding swap gates to make a 2-qubit gate (red) acting on distant qubits nearest-neighbor. (b) Adding new qubits so that the execution of gates are in a spatially local sequence.}
\label{fig:sparse-circuit}
\end{figure}

Second, we want to implement such an approximation of $U_{\rm PE}^\ideal$, where $\ket{E_\mu}$ is replaced with $\ket{\tilde{E}_\mu}$, using only a polynomial number of local gates.
Note that the phase estimation is applied to the unitary $e^{iHt}$, which is not a local unitary operation.
Nonetheless, for $O(1)$-local Hamiltonians, we can easily apply the standard Trotter decomposition~\cite{Trotter} to split $e^{iHt}$ into a product of local gates. 
We also want to use a discrete set of 2-qubit universal gates, which can be done by invoking the Solovay-Kitaev theorem~\cite{DawsonNielsen}.
This gives us $U_\PE^\local$, which will have some error $\zeta=\|(U_\PE^\local - U_{\rm PE}^\ideal) \ket{\psi}\ket{0^m} \|$.
This error $\zeta$ can be made small using only $O(\poly(n,\zeta^{-1}))$ resources.
Furthermore, this fact is true by definition for any digitally simulable Hamiltonian.
Then, as shown in Fig.~\ref{fig:sparse-circuit}(a), we make all gates to be nearest-neighbor on a 1D line by adding swap gates, obtaining $U_\PE^\NN$.
This fact is summarized as Proposition~\ref{prop:PEcircuit}, which we prove more formally in \cref{supp2}:
\begin{prop}
\label{prop:PEcircuit}
Given any $n$-qubit digitally simulable Hamiltonian $H=\sum_\mu E_\mu \ketbra{\psi_\mu}$, with a promise that the energies $E_\mu$'s lie in the interval $[0, E_{\max}]$, one can construct a circuit $U_\PE^\NN$ consisting of $\poly(n, \zeta^{-1})$ gates acting on $n+m$ qubits arranged on a 1D line with 1- or 2-qubit nearest-neighbor gates chosen from a discrete universal gate set, 
where $m=\poly(n)$, such that for any normalized input state $\sum_\mu c_\mu \ket{\psi_\mu}$
\begin{align}
    \bigg\|U_\PE^\NN \sum_\mu c_\mu \ket{\psi_\mu}\ket{0^m} - \sum_\mu c_\mu \ket{\psi_\mu}\ket{\tilde{E}_\mu}\bigg\| \le \zeta
\end{align}
where the first $s<m$ qubits of $\ket{\tilde{E}_\mu}$ can be read off as an $s$-bit approximation of the real number $\varphi_\mu = E_\mu/E_{\max}$.
\end{prop}

\subsection{Constructing 2D semi-TI strongly universal Hamiltonians}

To prove Theorem~\ref{thm:main}, we want to simulate any local (or digitally simulable) Hamiltonian $H$ with an $\cS$-Hamiltonian on a 2D square lattice with polynomial overhead, for any non-2SLD $\cS$. 
Based on the 1D phase estimation circuit produced by Proposition~\ref{prop:PEcircuit}, we construct a spatially sparse circuit Hamiltonian $H_\circuit$ that simulates our original target Hamiltonian $H$ (Proposition~\ref{prop:sparseHam}). Then the (semi-translation-invariant) $\cS$-Hamiltonian simulator on the 2D lattice is obtained from $H_\circuit$ by applying a series of gadgets. The full technical proof in given in \cref{supp3}, but we will sketch the essential ideas below.
To that end, we borrow
the notion of spatial sparsity from Ref.~\cite{OliveiraTerhal, UniversalHamiltonian} and generalize it to circuits:

\begin{defn}[Spatial sparsity of Hamiltonians and circuits (adapted from \cite{OliveiraTerhal})]
\label{defn:spatial-sparsity}
A Hamiltonian on $n$-qudits is \emph{spatially sparse} if its interaction hypergraph is one where (i) every vertex participates in $O(1)$ hyper-edges, and $(ii)$ there is a straight-line drawing in the plane such that every hyper-edge overlaps with $O(1)$ other hyper-edges, and the surface covered by every hyper-edge is $O(1)$.
Moreover, we say a quantum circuit $U=\prod_{t=1}^{T} U_t$ is \emph{spatially sparse} if there is a placement of the qudits in the two-dimensional plane such that (i) each qudit participates in $O(1)$ gates, and (ii) the spatial supports of $U_t$ and $U_{t+1}$ are only $O(1)$ distance apart for all $t$, each covering $O(1)$ contiguous area. 
\end{defn}

With the definition of spatial sparsity at hand, we now formally state the Proposition we want to show:

{
\setcounter{prop}{1}
\begin{prop}
\label{prop:sparseHam}
Given any $O(1)$-local $n$-qudit Hamiltonian $H$ with $\|H\|=O(\poly(n))$  and $d=O(1)$, one can construct a spatially sparse $5$-local Hamiltonian $H_\circuit$ that efficiently simulates $H$ (c.f. Definition~\ref{defn:CMPsimul}) to precision $(\Delta,\eta,\epsilon)$, with $\Delta=O(\epsilon^{-1}\|H\|^2 + \eta^{-1}\|H\|)$.
$H_\circuit$ has $O(\poly(n,\epsilon^{-1}))$ terms and qubits, and interaction energy at most $O(\poly(n,\eta^{-1},\epsilon^{-1}))$.
\end{prop}
}

To prove this Proposition, we start from the 1D phase estimation circuit $U_\PE^\NN$ obtained using Proposition~\ref{prop:PEcircuit} and construct a spatially sparse circuit using swap gates. 
To do this, we first make $U_\PE^\NN$ into a spatially sparse version, $U_{\rm PE}^\text{sparse}$, by moving it from the line to a 2D grid.
As visualized in Fig.~\ref{fig:sparse-circuit}(b), starting from the leftmost column of qubits, we apply just one nearest-neighbor gate before swapping all qubits to the next column and applying the next gate, and so on, getting us to  $U_{\rm PE}^{\rm sparse}$.
By ordering the gates in a snake-like fashion similar to Ref.~\cite{AharonovAQCUniversal, OliveiraTerhal}, we make sure that each qubit participates in only a constant number of gates, and that the temporally proximate gates in the sequence have spatially proximate support.

Before converting the circuit back to a Hamiltonian, we need to address the issue of the entanglement between each energy eigenstate and the ancilla register that has the energy bit-string, as evident in Eq.~\eqref{eq:PE-ideal-state}.
This incoherence between different eigenstates causes a large error for the simulation if not addressed.
We repair this error by running the circuit backwards (``uncomputing'') and then adding $L$ identity gates at the end (``idling''), so that most of the computational history of each eigenstate is simply the state itself.
We then get a new circuit  $U_\circuit = (\Id)^L (U_\PE^\sparse)^\dag (\Id)^s U_\PE^\sparse$, which is spatially sparse as long as $U_\PE^\sparse$ is. Note we inserted $s$ identity gates before applying $(U_\PE^\sparse)^\dag$ so that we can examine the energy bit-string bit-by-bit before it is uncomputed.
Although $U_\circuit=\Id$ is ultimately a trivial operation, the circuit's nontrivial temporal structure allows its computational history to encode the spectral information about the target Hamiltonian.

Subsequently, we apply Kitaev's circuit-to-Hamiltonian mapping~\cite{KSV02} to convert the spatially sparse, ``uncomputed" circuit $U_\circuit$ to a spatially sparse Hamiltonian $H_\circuit$. To do this in a spatially sparse way, we observe that in the circuit-to-Hamiltonian construction~\cite{KSV02}, each clock qubit corresponds to a single gate; then in our construction, that clock qubit is placed geometrically in the middle of the hyper-edge connecting the qubits participating in the gate, interacting with only nearby qubits.
Hence, the spatially sparsity of $H_\circuit$ is guaranteed by the spatial sparsity of $U_\circuit$.
The energy of the eigenstates of $H$ are restored by adding an appropriate bit-wise energy penalty on the $s$ qubits where the energy is written, while we idle between $U_\PE^\sparse$ and $(U_\PE^\sparse)^\dag$ [see Eq.~\eqref{eq:Hout-energy-penalty}].
We then use perturbative arguments (such as those shown in Ref.~\cite{BravyiHastingsSim}) to show that $H_{\circuit}$ simulates $H$ with polynomially small error, with only polynomial overheads.
This proves Proposition~\ref{prop:sparseHam}.

Finally, to prove Theorem~\ref{thm:main}, we map the spatially sparse Hamiltonian $H_\circuit$ to a Hamiltonian in a universal family on the 2D square lattice with additional polynomial overhead.
This is done via a sequence of reductions, with known techniques~\cite{OliveiraTerhal, UniversalHamiltonian}:
We first convert $H_\circuit$ to a real-valued Hamiltonian by doubling the number of qubits, and encoding any Pauli $Y$'s into a pair of $Y\otimes Y$.
We then remove all Pauli $Y$'s, and reduce the locality of the Hamiltonian to 2-local via applications of perturbative gadgets.
This is subsequently converted to a spatially sparse $\cS_0$-Hamiltonian where $\cS_0=\{XX+YY+ZZ\}$ or $\{XX+YY\}$.
Throughout this sequence of reductions, the spatial sparsity of the Hamiltonian is preserved, and both the interaction energy and qubit number only increases polynomially as the original system size and the target precision parameters $(\Delta, \eta, \epsilon)$.
This spatially sparse $\cS_0$-Hamiltonian involving only Pauli-interactions without $Y$'s can then be mapped to an $\cS_0$-Hamiltonian on the 2D lattice. Since the input Hamiltonian is spatially sparse, this mapping only incurs polynomial overhead in the interaction energy as shown in Ref.~\cite{OliveiraTerhal,UniversalHamiltonian}.
Furthermore, Ref.~\cite{UniversalHamiltonian} has shown that $\cS_0$-Hamiltonian on a 2D square lattice can be simulated by any $\cS$-Hamiltonian on a 2D square lattice for any non-2SLD $\cS$ (c.f. Definition~\ref{def:2SLD}), with polynomial overhead.
We have thus provided a construction that allows any such family of $\cS$-Hamiltonians on the 2D square lattice to efficiently simulate any local Hamiltonian with arbitrary geometry.

\vspace{-10pt}
\subsection{Constructing 1D universal Hamiltonians}
\vspace{-5pt}

We now describe how to construct a strongly universal family of Hamiltonian simulators in 1D.
To prove this result as stated in Theorem~\ref{thm:1D-universal}, we extend the circuit-to-1D-Hamiltonian constructions in Ref.~\cite{AharonovLine2009,Hallgren}.
These constructions were originally used to show QMA-completeness of Hamiltonians involving nearest-neighbor interaction on a 1D line of particles, by using them to encode the outcome of any circuit computation in their ground state energy.

The basic idea in these constructions is as follows.
Suppose we are given any quantum circuit consisting of $R$ rounds of nearest-neighbor gates on $n$ qubits in 1D such as $U_\PE^\NN$, where each round is of the form $U_{n-1,n}\cdots U_{23}U_{12}$ (some may be the identity gate).
We consider an equivalent circuit on a line of $2nR$ particles, divided into $R$ blocks, where each block encodes the computational state of the original $n$ qubits.
In this equivalent circuit, a single round of nearest-neighbor gate is applied in each block before the qubits are moved to the next block where subsequent gates can be performed.
The 8 internal states of each particle are necessary to store both the computational state of the original qubit and marker states that allows us to locally distinguish different stages of the computation of each particle (e.g., whether a gate has already been applied or needs to be applied).

We want to apply these constructions to simulate general Hamiltonians.
Following the same idea as in the proof of Theorem~\ref{thm:main}, we first convert the target Hamiltonian $H$ to a phase estimation circuit $U_\PE^\NN$ using nearest-neighbor gates on a line of qubits as in Proposition~\ref{prop:PEcircuit}.
Then, we use the scheme in Ref.~\cite{Hallgren} as well as our tricks of ``uncomputing'' and ``idling'' to map the circuit to a 1D Hamiltonian of nearest neighbor-interactions, and penalize the energy bit-string so that we simulate the full spectral properties of $H$.
However, a na\"ive application of this scheme would yield a highly nonlocal encoding, since the idling part of the circuit corresponds to simply moving the computational qubits down the line, causing the encoded eigenstates of $H$ to be delocalized over many blocks of particles.
To circumvent this issue, we modified the scheme so that the idling step is done without moving the computational qubits, while maintaining the consistency of all transition rules so that the legal computational history states are spectrally gapped from the rest.
Consequently, the eigenstates of $H$ are encoded in the qubit-subspace of some 8-state qudits from just one block of the line, yielding a local encoding. 
See \cref{supp4} for the full proof.

\section{Data Availability}
No datasets were generated or analyzed during the current study.

\section{Acknowledgements}
D. A. acknowledges the support of ISF grant No. 2137/19 and Advanced ERC grant 
No. 101201649 during the work on this research. 
L.Z. acknowledges funding from the Walter Burke Institute for Theoretical Physics at Caltech during the completion of this work.

\section{Author Contributions}
L.Z. and D.A. contributed to every aspect of the research and writing of this work.

\section{Competing Interests}
The authors declare no competing interests.


\bibliographystyle{naturemag}
\bibliography{UHrefs}

@article{Monroe2021ions,
  title = {Programmable quantum simulations of spin systems with trapped ions},
  author = {Monroe, C. and Campbell, W. C. and Duan, L.-M. and Gong, Z.-X. and Gorshkov, A. V. and Hess, P. W. and Islam, R. and Kim, K. and Linke, N. M. and Pagano, G. and Richerme, P. and Senko, C. and Yao, N. Y.},
  journal = {Rev. Mod. Phys.},
  volume = {93},
  issue = {2},
  pages = {025001},
  numpages = {57},
  year = {2021},
  month = {Apr},
  publisher = {American Physical Society},
  doi = {10.1103/RevModPhys.93.025001},
}

@article{Browaeys2020atoms,
   title={Many-body physics with individually controlled Rydberg atoms},
   volume={16},
   ISSN={1745-2481},
   DOI={10.1038/s41567-019-0733-z},
   number={2},
   journal={Nature Physics},
   publisher={Springer Science and Business Media LLC},
   author={Browaeys, Antoine and Lahaye, Thierry},
   year={2020},
   month=jan, pages={132–142} }

@article{Andersen2025thermalization,
  title = {Thermalization and criticality on an analogue–digital quantum simulator},
  volume = {638},
  ISSN = {1476-4687},
  DOI = {10.1038/s41586-024-08460-3},
  number = {8049},
  journal = {Nature},
  publisher = {Springer Science and Business Media LLC},
  author = {Andersen,  T. I. and Astrakhantsev,  N. and Karamlou,  A. H. and Berndtsson,  J. and Motruk,  J. and others},
  year = {2025},
  month = feb,
  pages = {79–85}
}

@article{Zhang2024observation,
  title = {Observation of microscopic confinement dynamics by a tunable topological $\theta$-angle},
  volume = {21},
  ISSN = {1745-2481},
  DOI = {10.1038/s41567-024-02702-x},
  number = {1},
  journal = {Nature Physics},
  publisher = {Springer Science and Business Media LLC},
  author = {Zhang,  Wei-Yong and Liu,  Ying and Cheng,  Yanting and He,  Ming-Gen and Wang,  Han-Yi and Wang,  Tian-Yi and Zhu,  Zi-Hang and Su,  Guo-Xian and Zhou,  Zhao-Yu and Zheng,  Yong-Guang and Sun,  Hui and Yang,  Bing and Hauke,  Philipp and Zheng,  Wei and Halimeh,  Jad C. and Yuan,  Zhen-Sheng and Pan,  Jian-Wei},
  year = {2024},
  month = dec,
  pages = {155–160}
}

@inproceedings{QSVT2019,
series={STOC ’19},
   title={Quantum singular value transformation and beyond: exponential improvements for quantum matrix arithmetics},
   url={http://dx.doi.org/10.1145/3313276.3316366},
   DOI={10.1145/3313276.3316366},
   booktitle={Proceedings of the 51st Annual ACM SIGACT Symposium on Theory of Computing},
   publisher={ACM},
   author={Gily\'{e}n, Andr\'{a}s and Su, Yuan and Low, Guang Hao and Wiebe, Nathan},
   year={2019},
   month=jun, pages={193–204},
   collection={STOC ’19} }

@article{Kohler2022general,
  title = {General Conditions for Universality of Quantum Hamiltonians},
  author = {Kohler, Tamara and Piddock, Stephen and Bausch, Johannes and Cubitt, Toby},
  journal = {PRX Quantum},
  volume = {3},
  issue = {1},
  pages = {010308},
  numpages = {14},
  year = {2022},
  month = {Jan},
  publisher = {American Physical Society},
  doi = {10.1103/PRXQuantum.3.010308},
  url = {https://link.aps.org/doi/10.1103/PRXQuantum.3.010308}
}

@article{Harley2024going,
  title = {Going beyond gadgets: the importance of scalability for analogue quantum simulators},
  volume = {15},
  ISSN = {2041-1723},
  DOI = {10.1038/s41467-024-50744-9},
  number = {1},
  journal = {Nature Communications},
  publisher = {Springer Science and Business Media LLC},
  author = {Harley,  Dylan and Datta,  Ishaun and Klausen,  Frederik Ravn and Bluhm,  Andreas and Fran\c{c}a,  Daniel Stilck and Werner,  Albert H. and Christandl,  Matthias},
  year = {2024},
  month = aug 
}

@article{DeutschQCT,
  doi = {10.1098/rspa.1985.0070},
  url = {https://doi.org/10.1098/rspa.1985.0070},
  year = {1985},
  publisher = {The Royal Society},
  volume = {400},
  number = {1818},
  pages = {97--117},
  title = {Quantum theory, the Church-Turing principle and the universal quantum computer},
  journal = {Proc. R. Soc. London Ser. A}
}

@book{KayeLaflammeMoscaBook,
author = {Kaye, Phillip and Laflamme, Raymond and Mosca, Michele},
title = {An Introduction to Quantum Computing},
year = {2007},
isbn = {0198570007},
publisher = {Oxford University Press, Inc.},
address = {USA}
}

@article{farhi2014quantum,
  title={{A Quantum Approximate Optimization Algorithm}},
  author={Farhi, Edward and Goldstone, Jeffrey and Gutmann, Sam},
  journal={arXiv preprint},
  eprint ={1411.4028},
  year={2014},
}

@article{MPQW,
author = {Andrew M. Childs  and David Gosset  and Zak Webb },
title = {Universal Computation by Multiparticle Quantum Walk},
journal = {Science},
volume = {339},
number = {6121},
pages = {791-794},
year = {2013},
doi = {10.1126/science.1229957},
URL = {https://www.science.org/doi/abs/10.1126/science.1229957},
}

@article{Ebadi2020,
	doi = {10.1038/s41586-021-03582-4},  
	year = 2021,
	volume = {595},
  
	number = {7866},
  
	pages = {227--232},
	author = {Sepehr Ebadi and Tout T. Wang and Harry Levine and Alexander Keesling and Giulia Semeghini and Ahmed Omran and Dolev Bluvstein and Rhine Samajdar and Hannes Pichler and Wen Wei Ho and Soonwon Choi and Subir Sachdev and Markus Greiner and Vladan Vuleti{\'{c}} and Mikhail D. Lukin},
	title = {Quantum phases of matter on a 256-atom programmable quantum simulator},
	journal = {Nature},
}

@article{Periwal2021,
	doi = {10.1038/s41586-021-04156-0},
	year = 2021,
	month = {dec},
	publisher = {Springer Science and Business Media {LLC}},
	volume = {600},
	number = {7890},
	pages = {630--635},
	author = {Avikar Periwal and Eric S. Cooper and Philipp Kunkel and Julian F. Wienand and Emily J. Davis and Monika Schleier-Smith},
	title = {Programmable interactions and emergent geometry in an array~of atom clouds},
	journal = {Nature}
}

@article{Kokail2019,
  doi = {10.1038/s41586-019-1177-4},
  url = {https://doi.org/10.1038/s41586-019-1177-4},
  year = {2019},
  month = may,
  publisher = {Springer Science and Business Media {LLC}},
  volume = {569},
  number = {7756},
  pages = {355--360},
  author = {C. Kokail and C. Maier and R. van Bijnen and T. Brydges and M. K. Joshi and P. Jurcevic and C. A. Muschik and P. Silvi and R. Blatt and C. F. Roos and P. Zoller},
  title = {Self-verifying variational quantum simulation of lattice models},
  journal = {Nature}
}

@article{Daley2022,
  doi = {10.1038/s41586-022-04940-6},
  url = {https://doi.org/10.1038/s41586-022-04940-6},
  year = {2022},
  month = jul,
  publisher = {Springer Science and Business Media {LLC}},
  volume = {607},
  number = {7920},
  pages = {667--676},
  author = {Andrew J. Daley and Immanuel Bloch and Christian Kokail and Stuart Flannigan and Natalie Pearson and Matthias Troyer and Peter Zoller},
  title = {Practical quantum advantage in quantum simulation},
  journal = {Nature}
}

@article{Choi2016,
archivePrefix = {arXiv},
arxivId = {1604.04178},
author = {Choi, Jae Yoon and Hild, Sebastian and Zeiher, Johannes and Schau{\ss}, Peter and Rubio-Abadal, Antonio and Yefsah, Tarik and Khemani, Vedika and Huse, David A. and Bloch, Immanuel and Gross, Christian},
doi = {10.1126/science.aaf8834},
eprint = {1604.04178},
issn = {10959203},
journal = {Science},
month = {jun},
number = {6293},
pages = {1547--1552},
publisher = {American Association for the Advancement of Science},
title = {{Exploring the many-body localization transition in two dimensions}},
url = {https://www.sciencemag.org/lookup/doi/10.1126/science.aaf8834},
volume = {352},
year = {2016}
}

@article{Trotter,
 ISSN = {00029939, 10886826},
 URL = {http://www.jstor.org/stable/2033649},
 author = {H. F. Trotter},
 journal = {Proceedings of the American Mathematical Society},
 number = {4},
 pages = {545--551},
 publisher = {American Mathematical Society},
 title = {On the Product of Semi-Groups of Operators},
 volume = {10},
 year = {1959}
}

@article{BB17,
	title = {{Universal Hamiltonians for Exponentially Long Simulation
}},
	author = {Thomas C. Bohdanowicz and Fernando G.S.L. Brand{\~a}o},
	arXivPrefix = "arXiv",
	eprint = {1710.02625},
	primaryClass = "quant-ph",
	year = 2017,
}

@article{UniversalHamiltonian,
	author = {Cubitt, Toby S. and Montanaro, Ashley and Piddock, Stephen},
	title = {{Universal quantum Hamiltonians}},
	volume = {115},
	number = {38},
	pages = {9497--9502},
	year = {2018},
	doi = {10.1073/pnas.1804949115},
	publisher = {National Academy of Sciences},
	issn = {0027-8424},
	journal = {Proceedings of the National Academy of Sciences}
}

@article{Piddock2020,
Author = {Stephen Piddock and Johannes Bausch},
Title = {{Universal Translationally-Invariant Hamiltonians}},
Year = {2020},
eprint = {arXiv:2001.08050},
}

@article{Kohler2020,
   title={Translationally Invariant Universal Quantum Hamiltonians in 1D},
   volume={23},
   ISSN={1424-0661},
   url={http://dx.doi.org/10.1007/s00023-021-01111-7},
   DOI={10.1007/s00023-021-01111-7},
   number={1},
   journal={Annales Henri Poincar\'{e}},
   publisher={Springer Science and Business Media LLC},
   author={Kohler, Tamara and Piddock, Stephen and Bausch, Johannes and Cubitt, Toby},
   year={2021},
   month=oct, pages={223–254} }

@article{AharonovLine2009,
  doi = {10.1007/s00220-008-0710-3},
  url = {https://doi.org/10.1007/s00220-008-0710-3},
  year = {2009},
  month = jan,
  publisher = {Springer Science and Business Media {LLC}},
  volume = {287},
  number = {1},
  pages = {41--65},
  author = {Dorit Aharonov and Daniel Gottesman and Sandy Irani and Julia Kempe},
  title = {The Power of Quantum Systems on a Line},
  journal = {Communications in Mathematical Physics}
}

@article{SYK1,
  title = {Gapless spin-fluid ground state in a random quantum Heisenberg magnet},
  author = {Sachdev, Subir and Ye, Jinwu},
  journal = {Phys. Rev. Lett.},
  volume = {70},
  issue = {21},
  pages = {3339--3342},
  numpages = {0},
  year = {1993},
  month = {May},
  publisher = {American Physical Society},
  doi = {10.1103/PhysRevLett.70.3339},
  url = {https://link.aps.org/doi/10.1103/PhysRevLett.70.3339}
}

@book{SYK2,
 author = {Alexei Kitaev}, 
 title = {A simple model of quantum holography},
 publisher = {KITP strings seminar and Entanglement 2015 program},
 url = {http://online.kitp.ucsb.edu/online/entangled15/},
 year = 2015,
}

@article{Hallgren,
 author = {Hallgren, Sean and Nagaj, Daniel and Narayanaswami, Sandeep},
 title = {{The Local Hamiltonian Problem on a Line with Eight States is QMA-complete}},
 journal = {Quantum Info. Comput.},
 issue_date = {September 2013},
 volume = {13},
 number = {9-10},
 month = sep,
 year = {2013},
 issn = {1533-7146},
 pages = {721--750},
 numpages = {30},
 acmid = {2535681},
 publisher = {Rinton Press, Incorporated},
 address = {Paramus, NJ},
}

@inproceedings{sparsify,
author = {Aharonov, Dorit and Zhou, Leo},
eprint = {1804.11084},
archivePrefix = {arXiv},
title = {{Hamiltonian Sparsification and Gap-Simulations}},
 booktitle = {Proceedings of the 2019 ACM Conference on Innovations in Theoretical Computer Science},
 series = {ITCS '19},
 year = {2019},
url = {http://arxiv.org/abs/1804.11084},
}

@article{Piddock2018Universal,
archivePrefix = {arXiv},
arxivId = {1802.07130},
author = {Piddock, Stephen and Montanaro, Ashley},
eprint = {1802.07130},
month = {feb},
title = {{Universal qudit Hamiltonians}},
url = {http://arxiv.org/abs/1802.07130},
year = {2018}
}

@article{AharonovAQCUniversal,
 author = {Aharonov, Dorit and van Dam, Wim and Kempe, Julia and Landau, Zeph and Lloyd, Seth and Regev, Oded},
 title = {Adiabatic Quantum Computation is Equivalent to Standard Quantum Computation},
 journal = {SIAM J. Comput.},
 issue_date = {April 2007},
 volume = {37},
 number = {1},
 month = Apr,
 year = {2007},
 issn = {0097-5397},
 pages = {166--194},
 numpages = {29},
 doi = {10.1137/S0097539705447323},
 acmid = {1328726},
 publisher = {Society for Industrial and Applied Mathematics},
 address = {Philadelphia, PA, USA},
}

@article{Feynman1982,
author = {Feynman, Richard P},
journal = {International Journal of Theoretical Physics},
number = {6},
title = {{Simulating Physics with Computers}},
volume = {21},
pages = {467--488},
doi = {10.1007/BF02650179},
year = {1982}
}

@article{BravyiHastingsSim,
  title = {On Complexity of the Quantum Ising Model},
  volume = {349},
  ISSN = {1432-0916},
  DOI = {10.1007/s00220-016-2787-4},
  number = {1},
  journal = {Commun. Math. Phys.},
  publisher = {Springer Science and Business Media LLC},
  author = {Bravyi,  Sergey and Hastings,  Matthew},
  year = {2016},
  month = nov,
  pages = {1–45}
}

@article{Preskill2018,
archivePrefix = {arXiv},
arxivId = {1801.00862},
author = {Preskill, John},
doi = {10.22331/q-2018-08-06-79},
eprint = {1801.00862},
issn = {2521-327X},
journal = {Quantum},
month = {Jan},
pages = {79},
publisher = {Verein zur F{\"{o}}rderung des Open Access Publizierens in den Quantenwissenschaften},
title = {{Quantum Computing in the NISQ era and beyond}},
url = {https://quantum-journal.org/papers/q-2018-08-06-79/},
volume = {2},
year = {2018}
}

@article{Bernien2017,
author = {Bernien, Hannes and Schwartz, Sylvain and Keesling, Alexander and Levine, Harry and Omran, Ahmed and Pichler, Hannes and Choi, Soonwon and Zibrov, Alexander S. and Endres, Manuel and Greiner, Markus and Vuleti{\'{c}}, Vladan and Lukin, Mikhail D.},
doi = {10.1038/nature24622},
issn = {0028-0836},
journal = {Nature},
month = {Nov},
number = {7682},
pages = {579--584},
publisher = {Nature Publishing Group},
title = {{Probing many-body dynamics on a 51-atom quantum simulator}},
url = {http://www.nature.com/doifinder/10.1038/nature24622},
volume = {551},
year = {2017}
}

@article{Zhang2017,
author = {Zhang, J. and Pagano, G. and Hess, P. W. and Kyprianidis, A. and Becker, P. and Kaplan, H. and Gorshkov, A. V. and Gong, Z.-X. and Monroe, C.},
doi = {10.1038/nature24654},
issn = {0028-0836},
journal = {Nature},
month = {Nov},
number = {7682},
pages = {601--604},
publisher = {Nature Publishing Group},
title = {{Observation of a many-body dynamical phase transition with a 53-qubit quantum simulator}},
url = {http://www.nature.com/doifinder/10.1038/nature24654},
volume = {551},
year = {2017}
}

@article{Arguello2019,
author = {Arg{\"{u}}ello-Luengo, Javier and Gonz{\'{a}}lez-Tudela, Alejandro and Shi, Tao and Zoller, Peter and Cirac, J Ignacio},
doi = {10.1038/s41586-019-1614-4},
issn = {1476-4687},
journal = {Nature},
number = {7777},
pages = {215--218},
title = {{Analogue quantum chemistry simulation}},
url = {https://doi.org/10.1038/s41586-019-1614-4},
volume = {574},
year = {2019}
}

@article{CiracZollerNatPhys2012,
	doi = {10.1038/nphys2275},
	Author = {Cirac, J. Ignacio and Zoller, Peter},
	Journal = {Nature Physics},
	M3 = {10.1038/nphys2275},
	Month = {Apr},
	Number = {4},
	Pages = {264--266},
	Title = {Goals and opportunities in quantum simulation},
	Url = {http://dx.doi.org/10.1038/nphys2275},
	Volume = {8},
	Year = {2012},
}

@article{KKR06,
    title = {{The Complexity of the Local Hamiltonian Problem}},
    author = {Julia Kempe and Alexei Kitaev and Oded Regev},
    journal = {SIAM J. Comput.},
    volume = 35,
    pages = {1070--1097},
    year = 2006,
}

@article{CaoNagajGadget,
    title = {{Perturbative gadgets without strong interactions}},
    author = {Yudong Cao and Daniel Nagaj},
    journal = {Quantum Inf. Comput.},
    volume = 15,
    pages = {1197--1222},
    year = 2015,
}

@article{OliveiraTerhal,
    title = {{The complexity of quantum spin systems on a two-dimensional square lattice}},
    author = {Roberto Oliveira and Barbara M. Terhal},
    journal = {Quantum Inf. Comput.},
    volume = 8,
    pages = {900--924},
    year = 2008,
}

@book{KSV02,
    author = {A. Yu Kitaev and A. Shen and M. N. Vyalyi},
    title =  {Classical and Quantum Computation},
    publisher = {American Mathematical Society},
    year = 2002,
}

@article{DawsonNielsen,
author = {Dawson, Christopher M. and Nielsen, Michael A.},
title = {{The Solovay-Kitaev Algorithm}},
year = {2006},
issue_date = {January 2006},
publisher = {Rinton Press, Incorporated},
address = {Paramus, NJ},
volume = {6},
number = {1},
issn = {1533-7146},
journal = {Quantum Info. Comput.},
month = jan,
pages = {81–95},
numpages = {15}
}

@article{FarhiAdiabatic2000,
archivePrefix = {arXiv},
arxivId = {quant-ph/0001106},
author = {Farhi, Edward and Goldstone, Jeffrey and Gutmann, Sam and Sipser, Michael},
eprint = {0001106},
month = {Jan},
primaryClass = {quant-ph},
title = {{Quantum Computation by Adiabatic Evolution}},
url = {http://arxiv.org/abs/quant-ph/0001106},
year = {2000}
}

@book{NielsenChuang,
 author = {Nielsen, Michael A. and Chuang, Isaac L.},
 title = {{Quantum Computation and Quantum Information}},
 year = {2011},
 edition = {10th},
 publisher = {Cambridge University Press},
 address = {New York, NY, USA},
}

\onecolumngrid

\appendix

\setcounter{secnumdepth}{2}

\section{Relationship between analog and dynamics simulations}
\label{supp1}
In this section, we prove Claim 1 in the main text about the relationship of analog and dynamics simulation.

We start by restating the definitions of local encodings and Hamiltonian simulations given in the main text, which will also be used in the rest of the appendices.

\begin{defn}[Local encoding, adapted from \cite{UniversalHamiltonian}]
\label{defn:local-encoding}
Consider an encoding map $\eps: \mathrm{Herm}((\CC^{d})^{\otimes n}) \to \mathrm{Herm}((\CC^{d'})^{\otimes n'})$ that takes Hermitian operators on $n$ qudits ($d$-dimensional particles) into operators acting on $n'$ $d'$-dimensional particles (where $d^n\le {d'}^{n'}$). 
We say $\eps$ is an \emph{$\eta$-local encoding} if we can write
\begin{equation} \label{eq:localenc}
\eps(A)=V(A\otimes P+\bar{A}\otimes Q)V^\dag,
\end{equation}
such that $V$ is an isometry satisfying $\|V-V_{\rm loc}\|\le \eta$ for some local isometry $V_{\rm loc}=\bigotimes_{i=1}^n V_i$, where each $V_i$ is an isometry acting on one qudit in the original system.
$P$ and $Q$ are locally orthogonal projectors, i.e., $P$ and $Q$ are orthogonal projectors and  $\forall i$ $\exists$  orthogonal projectors $P_i, Q_i$ acting on the same subsystem as $V_i$ such that $P_iQ_i=0$, $P_iP=P$ and $Q_iQ=Q$  (see also \cite[Supporting Information, Theorem 15]{UniversalHamiltonian}).
$\bar{A}$ is the complex conjugation of $A$. $\|\cdot\|$ is the spectral norm.

When $\eta=0$, we say $\eps$ is a \emph{local encoding}.

Furthermore, $\eps^\state$ is an $\eta$-local state-encoding if it is of the form $\eps^\state(\rho)=\eps'(\rho)/\Tr(\eps'(\rho))$ for some $\eta$-local encoding $\eps'$.
\end{defn}
\begin{defn}
[Analog Hamiltonian simulation, adapted from \cite{UniversalHamiltonian}]
\label{defn:CMPsimul}
A Hamiltonian $H'$ is a $(\Delta, \eta,\epsilon)$-\emph{simulation} of an $n$-qudit Hamiltonian $H$ if there exists an $\eta$-local encoding $\eps_\eta$ as in Definition~\ref{defn:local-encoding}  such that
\begin{enumerate}
\item 
$\eps_\eta(\Id) = P_{\le \Delta(H')}$; 
\item $\|H'_{\le \Delta} - \eps_\eta(H)\| \le \epsilon$.
\end{enumerate}
Here, $P_{\le \Delta(H')}$ is the projector onto the subspace of eigenstates of $H'$ with eigenvalue $\le \Delta$, and
$H'_{\le \Delta} =  P_{\le \Delta(H')} H'$ is the restriction of $H'$ onto these states. 
We say the simulation is \emph{efficient} if both the number of particles in $H'$ and its maximum energy $\|H'\|$ are at most $O(\poly(n,\eta^{-1},\epsilon^{-1}, \Delta))$,  
and the description of $H'$ is efficiently computable.
\end{defn}

\begin{defn}[dynamics simulation]
\label{defn:dynamics}
A Hamiltonian $H'$ is an \emph{$(\eta, \epsilon)$-dynamics simulation} of an $n$-qudit Hamiltonian $H$ if for any time $t\ge 0$, there exist (possibly time-dependent) encoding maps $\eps_{\rm in}^\state(\rho,t)$ and $\eps_{\rm out}^\state(\rho, t)$, which are $\eta$-local state-encodings as in Definition~\ref{defn:local-encoding}, such that for any density matrix $\rho_0$, we have 
\begin{equation}
\label{eq:dynamics}
\left\|e^{-iH't'}\eps_{\rm in}^\state(\rho_0, t) e^{iH' t'} - \eps_{\rm out}^\state(e^{-iHt}\rho_0 e^{iHt}, t)\right\|_1 \le \epsilon t,
\end{equation}
where $t' = \poly(t,n)$.
\end{defn}

We now restate Claim 1 from the main text, which we prove in the remainder of this section:
\begin{claim}
\label{claim:dyn-vs-analog}
If $H'$ is a $(\Delta,\eta,\epsilon)$-analog simulation of $H$ as in Definition~\ref{defn:CMPsimul}, then $H'$ is an $(\eta,2\epsilon)$-dynamics simulation of $H$.
Conversely, if $H'$ is an $(\eta, 0)$-dynamics simulation of $H$ with time-independent encodings $\eps_{\rm in}^\state(\rho,t)=\eps_{\rm out}^\state(\rho,t)=V\rho V^\dag$ and $t'=t$, then $H'+c I$ restricted to a certain subspace is a $(\Delta, \eta, 0)$-analog simulation of $H$, for some offset $c$ and some $\Delta \ge \|H\|$.
\end{claim}

\begin{proof}
First we show the easier direction: a good analog simulator is a good dynamics simulator.
This follows somewhat straightforwardly from the properties of analog simulation (see also \cite[Supporting Information, Corollary 29]{UniversalHamiltonian}). 
Given that $H'$ is a $(\Delta,\eta,\epsilon)$-simulation of $H$, then by Definition~\ref{defn:CMPsimul} there is an $\eta$-local encoding $\eps_\eta(A) = V(A\otimes P + \bar{A}\otimes Q)V^\dag$ such that $\eps_\eta(\Id) = P_{\le \Delta(H')}$ and $\|H'_{\le \Delta} - \eps_\eta(H)\| \le \epsilon$. 

\begin{equation}
e^{-i\eps_\eta(H)t} = V \exp[{-i(H\otimes P + \bar{H}\otimes Q)t}] V^\dag = V(e^{-iHt}\otimes P + e^{-i\bar{H}t}\otimes Q) V^\dag,
\end{equation}
where the last equality follows from the fact that $P$ and $Q$ are orthogonal projectors and $PQ=0$.

Suppose for now $\rank(P)\ge 1$. Let $\sigma$ be any state where $P\sigma=\sigma$.
Let $\eps^\state(\rho) = V(\rho\otimes \sigma)V^\dag$, which is an $\eta$-local state-encoding since $\eps_\eta$ is an $\eta$-local encoding.
Then 
\begin{equation}
\eps^\state(e^{-iHt}\rho_0 e^{iHt}) = V(e^{-iHt}\rho_0 e^{iHt}\otimes \sigma) V^\dag = e^{-i\eps_\eta(H)t}\eps^\state(\rho_0)e^{i\eps_\eta(H)t},
\end{equation}
where the last equality uses the fact that $Q\sigma=0$.

Applying these identities, we have
\begin{align}
\left\|e^{-iH't}\eps^\state(\rho_0) e^{iH' t} - \eps^\state(e^{-iHt}\rho_0 e^{iHt})\right\|_1
&= \left\|e^{-iH't}\eps^\state(\rho_0) e^{iH' t} - e^{-i\eps_\eta(H)t}\eps^\state(\rho_0)e^{i\eps_\eta(H)t}\right\|_1 \nonumber \\
&= \left\|e^{-iH'_{\le \Delta} t}\eps^\state(\rho_0) e^{iH'_{\le\Delta} t} - e^{-i\eps_\eta(H)t}\eps^\state(\rho_0)e^{i\eps_\eta(H)t}\right\|_1 \nonumber \\
&\le 2 \left\|e^{-iH'_{\le \Delta} t} - e^{-i\eps_\eta(H)t}\right\| \le 2 \left\|H'_{\le \Delta} - \eps_\eta(H)\right\|t \le 2\epsilon t.
\end{align}
The second line uses the fact that $\eps^\state(\rho_0)$ is in the subspace $P_{\le \Delta(H')}$.
In the last line, the first inequality comes from the fact that $\|ACA^\dag - BCB^\dag\|_1 \le (\|A\|+\|B\|) \|A-B\| \|C\|_1$ (see \cite[Supporting Information, Lemma 17]{UniversalHamiltonian}).
The second inequality follows from the fact that $\partial_t[e^{iAt}e^{-iBt}-\Id] = e^{iAt}(A-B)e^{-iBt}$.
Letting $\eps_{\rm in}^\state(\cdot, t) =\eps_{\rm out}^\state(\cdot,t)=\eps^\state(\cdot)$ and $t'=t$, this shows that $H'$ is an $(\eta,2\epsilon)$-dynamics simulation of $H$ as in Definition~\ref{defn:dynamics}.

For the other case of $\rank(P)=0$, we necessarily have $\rank(Q)\ge 1$. Let $\sigma$ be any state where $Q\sigma=\sigma$, and $\eps^\state(\rho) = V(\bar{\rho}\otimes \sigma)V^\dag$. Then $\eps^\state(e^{-iHt}\rho_0 e^{iHt}) = e^{i\eps_\eta(H)t}\eps^\state(\rho_0)e^{-i\eps_\eta(H)t}$, which is a time-reversed simulation. Hence, letting $t'=-t$, the same argument as above shows that $H'$ is an $(\eta,2\epsilon)$-dynamics simulation of $H$.

Now we show that the converse direction is almost true.
Given that $H'$ is an $(\eta,0)$-dynamics simulation of $H$ with time-independent encodings $\eps_{\rm in}^\state(\rho,t)=\eps_{\rm out}^\state(\rho,t)=V\rho V^\dag$ and $t'=t$.
Then Definition~\ref{defn:dynamics} implies that for any input state $\rho_0$ and time $t\ge 0$, 
\begin{equation}
e^{-iH't} V\rho_0 V^\dag e^{iH't}  = Ve^{-iH t} \rho_0 e^{iHt} V^\dag.
\end{equation}
Let us write $H=\sum_\mu E_\mu \ketbra{\psi_\mu}$ in its eigendecomposition.
Consider $\rho_0=\ketbra{\psi_\mu}$ as input state, then
\begin{equation}
e^{-iH't} V\ketbra{\psi_\mu}V^\dag e^{iH't} = V\ketbra{\psi_\mu}V^\dag,
\end{equation}
which implies that $\ket{\psi_\mu'} := V\ket{\psi_\mu}$ is an eigenstate of $H'$.
Let us denote $E'_\mu$ as the (currently unknown) corresponding eigenvalue.
Next, for any two eigenstate $\ket{\psi_1}$ and $\ket{\psi_2}$ of $H$, consider $(\ket{\psi_1} + \ket{\psi_2})/\sqrt{2}$ as input state.
Then
\begin{equation}
Ve^{-iHt} \rho_0 e^{iHt}V^\dag = \frac12(\ketbra{\psi'_1}+ \ketbra{\psi'_2}  +  e^{i\delta_{12}t}\ketbrat{\psi'_1}{\psi'_2} + e^{-i\delta_{12}t}\ketbrat{\psi'_2}{\psi'_1}),
\end{equation}
where we have denoted $\delta_{12} = E_2-E_1$. Let $\delta'_{12} = E'_2- E'_1$, then
\begin{align}
e^{-iH't} V\rho_0 V^\dag e^{iH't} = \frac12(\ketbra{\psi'_1} + \ketbra{\psi'_2} + e^{i\delta'_{12}t} \ketbrat{\psi'_1}{\psi'_2} + e^{-i\delta'_{12} t}\ketbrat{\psi'_2}{\psi'_1}).
\end{align}
Since these two are equal for any $t\ge 0$, we must have $\delta'_{12} = \delta_{12}$.
Hence, $H'$ preserves the relative difference between any two eigenvalues of $H$, but there could be an arbitrary offset between the two set of eigenvalues $\{E_\mu\}_\mu$ and $\{E'_\mu\}_\mu$.
Let $E_0=\min_\mu E_\mu$ and $E'_0 = \min_\mu E'_\mu$.
Let $\L=\spn\{\ket{\psi'_\mu}\}_\mu$ be the subspace spanned by the encoded eigenstates of $H$, possibly augmented by excited states of $H'$ with energy $\ge \max_\mu E'_\mu$.
Then $H'+(E_0-E_0')\Id$ restricted to $\L$ is a $(\Delta,\eta,0)$-simulation of $H$ for some $\Delta\ge \|H\|$, with $\eta$-local encoding $\eps_\eta(A)=VAV^\dag$.
\end{proof}

\section{A 1D nearest-neighbor implementation of phase estimation circuit\label{sec:1D-PE}}
\label{supp2}
In this section, we formally show that given any digitally simulable Hamiltonian $H$ (including all local Hamiltonians), one can construct a phase estimation circuit such that the energy of any input eigenstate of $H$ can be written down as bits on some ancilla qubits to $O(\log n)$ bit precision with $O(1/\poly(n))$ error in the state.
The proof uses standard techniques such as Trotter decomposition and Solovay-Kitaev algorithm.
In particular, we show that this can be done with a circuit acting on a line of qubits with nearest-neighbor gates.
This will serve as the first step of our efficient construction of universal Hamiltonian simulators.

\begin{prop}[formal]
\label{prop:PEcircuit}
Consider any digitally simulable Hamiltonian $H=\sum_{a} H_a = \sum_\mu E_\mu \ketbra{\psi_\mu}$ acting on $n$ qubits, where we assume w.l.o.g. that $0 \le E_\mu \le E_{\max}$ for some known number $E_{\max}= O(\poly(n))$.
For any $s = O(\log n)$ and $\zeta>0$, we can construct a phase estimation circuit $U_\PE^\NN$ consisting of $O(\poly(n,\zeta^{-1}))$ 1- or 2-qubit nearest neighbor gates drawn from any universal gate set, acting on a line of $n+m$ qubits, where $m=O(\poly (n))$ such that the following is true:
For any normalized state $\sum_\mu c_\mu \ket{\psi_\mu}$, the circuit $U_\PE^\NN$ satisfies
\begin{align}
\left\| U_\PE^\NN \sum_\mu c_\mu \ket{\psi_\mu}\ket{0^m} - \sum_\mu c_\mu \ket{\psi_\mu}\ket{\tilde{E}_\mu}\ket{\rest_\mu}\right\| \le  \zeta
\end{align}
where $\ket{\tilde{E}_\mu}=\ket{\varphi_{\mu,1}\varphi_{\mu,2}\varphi_{\mu,3}\cdots\varphi_{\mu,s}}$ is the $s$-bit truncated representation of $\varphi_\mu = E_\mu/E_{\max} = 0.\varphi_{\mu,1}\varphi_{\mu,2}\varphi_{\mu,3}\cdots$  with $\varphi_{\mu,j}\in \{0,1\}$,
and $\ket{\rest_\mu}$ is some unimportant, normalized state on the remaining $m-s$ ancilla qubits.
\end{prop}

\begin{proof}
Let us first consider the standard implementation of quantum phase estimation algorithm circuit $U_{\PE}$.
Here, the circuit uses the evolution operator $u_j=e^{iH\tau 2^{j-1}}$ under $H$, where $\tau = 2\pi/E_{\max}$, and writes phase of the eigenvalues of $u_1=e^{iH\tau}$ on some ancilla qubits.
Note the eigenvalues of $u_1$ are $e^{i2\pi\varphi_\mu}$, where $\varphi_\mu=E_\mu \tau/(2\pi)$.
Since $0 \le \varphi_\mu\le 1$, we can write $\varphi_\mu = 0.\varphi_{\mu,1}\varphi_{\mu,2}\varphi_{\mu,3}\ldots$, where $\varphi_{\mu,j} \in \{0,1\}$ as its binary representation.

Ideally, the action of the phase estimation circuit on input states $\{\ket{\psi_\mu}\ket{0^m}\}_{\mu=1}^{2^n}$ is
\begin{equation}
\label{eq:ideal-PE}
U_{\rm PE}^\ideal \ket{\psi_\mu}\ket{0^m} = \ket{\psi_\mu}\ket{E_\mu},
\end{equation}
where $\ket{E_\mu}=\ket{\varphi_{\mu,1}}\otimes \ket{\varphi_{\mu,2}}\otimes\cdots \otimes \ket{\varphi_{\mu,m}}$ gives a $m$-bit string that exactly represents the real number $E_\mu/E_{\max}$.

In reality, there are two sources of errors that cause our desired 1D implementation of the phase estimation circuit to deviate from $U_{\rm PE}^\ideal$.

\textbf{Error 1: finite bit-precision.}---
The first  error is due to the fact that the energy eigenvalues don't generally have finite-bit-precision representation.
Let us denote $\tilde{E}_\mu=2\pi\tilde{\varphi}_\mu/\tau$ as an approximate value of the energy $E_\mu$ with $s$-bit precision, where $\tilde{\varphi}_\mu = 0.\varphi_{\mu,1}\varphi_{\mu,2}\varphi_{\mu,3}\cdots\varphi_{\mu,s}$.

The error comes from the fact that generally, $|E_\mu-\tilde{E}_\mu|=O(2^{-s})$ is non-zero.
In other words, since $\varphi_\mu \neq \tilde{\varphi}_\mu$, there's  error from imprecise phase estimation.
Let us consider a phase estimation circuit $U_{\rm PE}$ implemented to $p$-bit precision, where $p > s$.
Let $b_\mu$ be the integer in the range $[0,2^{p}-1]$ such that $0\le \varphi_\mu - b_\mu/2^p \le 2^{-p}$.
It is well-known~\cite{NielsenChuang} that the action of the standard implementation of $U_{\rm PE}$ on any input state $\ket{\psi_\mu}\ket{0}$ result in the following state
\begin{align}
U_{\rm PE} \ket{\psi_\mu}\ket{0^m} &= \ket{\psi_\mu} \otimes \bigg( \frac{1}{2^{p}}\sum_{k,\ell=0}^{2^p-1} e^{-i2\pi  k \ell/2^p} e^{i 2\pi \varphi_\mu k} \ket{\ell} \bigg) \otimes \ket{\extra_\mu}= \ket{\psi_\mu} \otimes \Big(\sum_{\ell=0}^{2^p-1}\alpha_\ell^\mu \ket{\ell}\Big) \otimes \ket{\extra_\mu}
\end{align}
where $\ket{\ell}=\ket{\ell_1\cdots \ell_p}$ is the binary representation of $\ell$,
\begin{align}
\alpha_\ell^\mu &= \frac{1}{2^p}\sum_{k=0}^{2^p-1} \left[e^{i 2\pi (\varphi_\mu-\ell/2^p)}\right]^k = \frac{1}{2^p} \left[\frac{1-e^{i 2\pi (2^{s}\varphi_\mu-\ell)}}{1-e^{i 2\pi (\varphi_\mu-\ell/2^p)}} \right],
\end{align}
and $\ket{\extra_\mu}$ is some normalized state on the remaining $m-p$ ancilla qubits that may be storing some unimportant information about the computation upon input $\ket{\psi_\mu}\ket{0^m}$.
The analysis from Sec.~5.2.1 in Ref.~\cite{NielsenChuang} shows that the probability of getting a state that is a distance of $e$ integer away is
\begin{equation}
p_\mu^\text{error}(e) \equiv \sum_{|\ell-b_\mu| > e} |\alpha_\ell^\mu|^2 \le \frac{1}{2(e-1)}
\end{equation}
Note that we only care about the first $s<p$ bits, so we can choose $e=2^{p-s}-1$.
Hence,
\begin{align}
U_{\rm PE} \ket{\psi_\mu}\ket{0^m} &=  \ket{\psi_\mu} \otimes \left[ \sum_{|\ell-b_\mu|\le e} \alpha_\ell^\mu \ket{\ell}  + \sum_{|\ell-b_\mu|> e} \alpha_\ell^\mu \ket{\ell} \right] \otimes \ket{\extra_\mu} \nonumber \\
&= \ket{\psi_\mu} \otimes\left(\sqrt{1-p_\mu^\text{error}}\ket{\tilde{E}_\mu}\ket{\rest^1_\mu} + \sqrt{p_\mu^\text{error}}\ket{\rest^2_\mu}\right)
\otimes \ket{\extra_\mu}.
\end{align}
Here, $\ket{\tilde{E}_\mu}=\ket{\varphi_{\mu,1}\varphi_{\mu,2}\varphi_{\mu,3}\cdots\varphi_{\mu,s}}$ represents the first $s$ bits of $\varphi_\mu$, $\ket{\rest_\mu^1}$ is a normalized superposition over the remaining $p-s$ qubit coming from the bitstrings which agree with $\varphi_\mu\approx b_\mu/2^p$ on the first $s$ bits, and $\ket{\rest_\mu^2}$ is a normalized superposition of $p$-bit strings whose values deviate from the good estimate $b_\mu$ of $\varphi_\mu$ by more than $e$.
Comparing this with the idealized output in Eq.~\eqref{eq:ideal-PE}, we can identify $\ket{\rest_\mu}:=\ket{\rest_\mu^1}\otimes \ket{\extra_\mu}$, and observe that
\begin{align}
(U_{\rm PE} - U_{\rm PE}^\ideal) \ket{\psi_\mu}\ket{0^m} = \ket{\psi_\mu}\ket{\text{error}_\mu}, \quad
\text{where}
\quad
\left\|\ket{\text{error}_\mu}\right\|^2 \le 2p^\text{error}_\mu = O(2^{-(p-s)})
\end{align}
Thus, for any normalized state $\ket{\psi}=\sum_\mu c_\mu \ket{\psi_\mu}\ket{0^m}$, we have
\begin{align}
\|(U_{\rm PE} - U_{\rm PE}^\ideal)\sum_\mu c_\mu \ket{\psi_\mu}\ket{0^m}\|^2 = O(2^{-(p-s)}) \le {\zeta/2}
\end{align}
where we choose, for example, $p=2s + O(\log \zeta^{-1})$, and thus make this first source of error due to finite bit-precision to be smaller than $\zeta/2$ for any given $\zeta>0$.

\textbf{Error 2: local gate approximation of $e^{-iH\tau_j}$.}---
The second source of error is due to the fact that we need to implement the circuit $U_\PE$ using only 1 or 2-qubit gates, in order to ensure the corresponding circuit-Hamiltonian is local,
The only non-local gates in the $p$-bit precise phase estimation algorithm that we need to address are the controlled-application of Hamiltonian evolution, $\ketbra{0} \otimes \Id + \ketbra{1}\otimes u_j$, where
$u_j=e^{-iH\tau_j}$, $\tau_j=2^{j-1}\tau=2^j\pi/E_{\max}$ and $j=1,2,\ldots,p$.

We address this by invoking the fact that $H$ is digitally simulable, which by definition enables one to construct a quantum circuit $W$ with $\poly(n,\tau,\xi^{-1})$ one- or two-qubit gates drawn from a universal set such that $\|W-e^{-iH\tau}\| \le \xi$.
Choosing $\xi=\zeta/(4p)$ and applying this construction to $\{u_j\}_{j=1}^p$, this yield a phase-estimation circuit containing $\poly(n,\zeta^{-1})$ local gates acting on at most three qubits each. Further reduction to one- and two-qubit universal gates can be performed by using the Solovay-Kitaev algorithm~\cite{DawsonNielsen}, incurring $\poly\log\zeta$ overhead.
The final circuit is $U_\PE^\local$ consisting of only $R_\local = \poly(n,\zeta^{-1})$ one- or two-qubit gates, satisfying
\begin{equation}
    \|U_\PE^\local - U_\PE\| \le \zeta/2.
\end{equation}

Lastly, we ensure that this circuit $U_\PE^{\rm local}$ only consists of nearest-neighbor gates by the following procedure:
\begin{enumerate}
\item Place all $n$ qubits on a line with any pre-determined ordering.
\item Iterate over each gate $U_t$, $t=1,\ldots R_\local$. If $U_t$ acts on qubits that are not neighbors on the line, add a sequence $S_t$ of swap gates on nearest neighbors in the circuit before $U_t$ so that $U_t$ acts on neighbors. Then add the same swap gates in reversed order in the circuit after $U_t$ so that the qubits returned to their original order on the line.
At the end of this step, the new circuit is of the form $U_{\rm PE}^{\rm NN} = \prod_{t=1}^{R_0} (S_t^\dag U_t S_t)$, with $R_0=O(R_\local N)$ gates, each only acting on a neighbors group of qubits. See Fig.~3(a) in the main text for an example.
\end{enumerate}
Note $U_\PE^\NN$ is effectively equivalent to $U_\PE^\local$ since we are simply swapping qubits around.

Putting everything together, we have $U_\PE^\NN \equiv U_\PE^\local \approx U_\PE \approx U_\PE^\ideal$, where each of the approximate equality has error bounded by $\zeta/2$.
In conclusion, for any $\zeta>0$, we can construct a phase estimation circuit $U_\PE^\NN$ comprised of only $O(\poly(n,\zeta^{-1}))$ 1-qubit or 2-qubit nearest-neighbor gates on a line, such that its action is $\zeta$-close to $U_{\rm PE}^\ideal$ on any normalized input state $\ket{\psi} = \sum_\mu c_\mu \ket{\psi_\mu}\ket{0^m}$:
\begin{align}
\|(U_{\rm PE}^\NN - U_{\rm PE}^\ideal) \ket{\psi}\ket{0^m} \| \le \zeta .
\label{eq:PE-error}
\end{align}
\end{proof}

\section{Proof that spin models on 2D lattice are strongly universal\label{sec:2D-proof}}
\label{supp3}

In this section, we show the details of the construction for strongly universal Hamiltonians on a 2D square lattice, proving the following theorem:

\begin{thm}
\label{thm:main}
Any $\S$-Hamiltonian family on the 2D square lattice is strongly universal, as long as $\S$ is non-2SLD.
In particular, it's sufficient for $\S$ to contain only a single interaction (such as Heisenberg or XY-interaction), implying that there are semi-translation-invariant Hamiltonians in 2D that are strongly universal. 
\end{thm}

\subsection{Efficient and spatially sparse Hamiltonian simulator\label{sec:sparse-sim}}

Before proving the strong universality of 2D spin-lattice models, we first prove the following result, where we show that any local Hamiltonian can be simulated by a spatially sparse Hamiltonian.

\begin{prop}
\label{prop:sparseHam}
Given any $O(1)$-local $n$-qudit Hamiltonian $H$ with $\|H\|=O(\poly(n))$ and $d=O(1)$, one can construct a spatially sparse $5$-local Hamiltonian $H_\circuit$ that efficiently simulates $H$ to precision $(\Delta,\eta,\epsilon)$, with $\Delta=O(\epsilon^{-1}\|H\|^2 + \eta^{-1}\|H\|)$.
$H_\circuit$ has $O(\poly(n, \epsilon^{-1}))$ terms and qubits, and interaction energy at most $O(\poly(n,\eta^{-1},\epsilon^{-1}))$.
\end{prop}

To prove the above Proposition, we first prove two smaller results in Lemmas~\ref{lem:circuit-ham-simul} and \ref{lem:circuit-idling} about different aspects of using the Feynman-Kitaev circuit-to-Hamiltonian construction~\cite{KSV02} for analog Hamiltonian simulation.
The following concept of history states will be useful in our discussion:

\begin{defn}[history state]
Let $U=U_T \cdots U_2 U_1$ be a quantum circuit acting on $n+m$ qudits.
Then for any input state $\ket{\psi_\mu}\in \CC^{d^n}$ and $m=\poly(n)$, the \emph{history state} with respect to $U$ and $\ket{\psi_\mu}$ is the following
\begin{equation}
\ket{\eta_\mu} = \frac{1}{\sqrt{T+1}}\sum_{t=0}^T \Big( U_{t}\cdots U_2 U_1 \ket{\psi_\mu}\ket{0^m}^\anc \Big)\ket{1^t 0^{T-t}}^\clock
\end{equation}
\end{defn}

We now state the first of the two lemmas, which describes a circuit-to-Hamiltonian transformation that can be used for analog Hamiltonian simulation, assuming an appropriate energy penalty Hamiltonian $H_{out}$ can be constructed.

\begin{lemma}[Circuit-Hamiltonian simulation]
\label{lem:circuit-ham-simul}
Let $H$ be any target $n$-qu$d$it Hamiltonian, with an orthonormal basis for eigenstates given by $\{\ket{\psi_\mu}\}_{\mu=1}^{d^n}$.
Let $U=\prod_{t=1}^T U_t$ be a quantum circuit with $T$ gates, each at most $k$-local.
Let $\L = \spn\{\ket{\eta_\mu}\}_{\mu=1}^{d^n}$ be the subspace of history states with respect to $U$ and $\{\ket{\psi_\mu}\}$.
Suppose there exists a Hamiltonian $H_{out}$ such that
\begin{align}
    \|V H V^\dag - H_{out}|_{\L}\| \le \epsilon/2
\end{align}
where $\eps(H) = V HV^\dag$ is a local encoding, 
and $O|_{\L}$ means operator $O$ restricted to subspace $\L$.
Then for any $\eta>0$, we can construct a Hamiltonian $H_\circuit$ from the description of $U$ and $H_{out}$ such
that $H_\circuit$ is a $(\Delta, \eta,\epsilon)$-simulation of $H$ with local encoding $\eps$, where $\Delta\ge O (\epsilon^{-1}\|H_{out}\|^2 + \eta^{-1}\|H_{out}\|)$, as in Definition~\ref{defn:CMPsimul}.
The constructed ${H}_\circuit$ is $(k+3)$-local, has $O(T)$ terms and particles, and uses $O(\poly(n,T,\Delta))$ interaction energy.
Furthermore, $H_\circuit$ is spatially sparse if the circuit $U$ is spatially sparse.
\end{lemma}

The second lemma concerns an ``idling'' trick, where we transform a circuit so that its corresponding history states are modified to increase the simulation precision under a desired local encoding $\eps(H) = H\otimes \ketbra{\alpha}$ (i.e. $V$ is the isometry acting as $V\ket{\psi}=\ket{\psi}\ket{\alpha}$).

\begin{lemma}[Idling to enhance simulation precision]
\label{lem:circuit-idling}
Consider an uncomputed quantum circuit $U_{D}\cdots U_1=\Id$.
Suppose we add $L$ identity gates to the end of the circuit, so that we obtain a new circuit $U=\Id^L U_D\cdots U_1$ with length $T=D+L$.
Let $H=\sum_\mu E_\mu \ketbra{\psi_\mu}$ be any Hamiltonian. Furthermore, let $\ket{\eta_\mu}$ be the history state with respect to $U$ and $\ket{\psi_\mu}$, and suppose $H_\eff = \sum_\mu E_\mu\ketbra{\eta_\mu}$.
For any $\epsilon > 0$, if we choose $L=O(\frac{D\|H\|^2}{\epsilon^2})$, then there is an ancilla state $\ket{\alpha}$ such that $\|H\otimes \ketbra{\alpha} - H_\eff \| \le \epsilon$.
\end{lemma}

With the two lemmas stated, we can now prove our main Proposition~\ref{prop:sparseHam}:

\begin{proof}[\textbf{Proof of Proposition~\ref{prop:sparseHam}}]
Given any $O(1)$-local $n$-qu$d$it Hamiltonian where $d$ is a constant, we can easily convert it to an $O(1)$-local $O(n)$-qubit Hamiltonian by simply encoding each qu$d$it in the subspace of a group of $\lceil\log_2 d\rceil$ qubits.
We can separate the extra states in this redundant encoding (when $d$ is not a power of 2) from the relevant part of spectrum by adding to the Hamiltonian a local energy penalty term on acting each group with $\|H\|=O(\poly(n))$ magnitude.
Hence, we will call $H$ the $O(1)$-local $n$-qubit Hamiltonian containing $O(\poly (n))$-strength interactions obtained after this conversion.

Let us denote the normalized eigenstates of $H$ as $\ket{\psi_\mu}$, with corresponding eigenvalues $E_\mu$.
We assume they are ordered such that $E_1  \le E_2 \le E_3 \le \cdots \le E_{2^n}$.

From Proposition~\ref{prop:PEcircuit}, for any $s=O(\log n)$, we can construct a $\zeta$-approximate, $s$-bit precise phase estimation circuit $U_\PE^\NN$ such that it acts on a line of $N = O(\poly(n))$ qubits with $R_0 = O(\poly(n,\zeta^{-1}))$ nearest-neighbor gates.
We want to replace it with a spatially sparse circuit $U_\PE^\text{sparse}$ with $O(R_0 N)$ qubits and gates (see Definition~8 in the main text).
This can be done with polynomial overhead in the same way as done in Ref.~\cite{OliveiraTerhal, AharonovAQCUniversal}.
We begin by placing the $N$ original qubits on the first column of a $N\times R_0$ grid of qubits. For column $i=1,2,\ldots,R_0$, we execute only the $i$-th gate from the circuit $U_\PE^\NN$, and other uninvolved qubits are acted on by identity gates. After each column, we swap the state of the qubit of column $i$ to $i+1$.
We order the execution of all the gates such that the gates from $U_\PE^\NN$ and identity gates are executed top-to-bottom, and the swap gates between column are executed from bottom-to-up [see Fig.~3(b) in the main text].
It is thus easy to see that in this new circuit $U_\PE^\text{sparse}$, each qubit participates in at most 3 gates (up to two swap gates and a non-trivial gate), and the gate are executed in a spatially local sequence.
Note the action of $U_\PE^\text{sparse}$ is equivalent to $U_\PE^\NN$ up to re-ordering of the qubits, since we've only added swap gates. Thus, Proposition~\ref{prop:PEcircuit} gives us
\begin{align} \label{eq:PE-approx}
\left\| U_\PE^\text{sparse} \sum_\mu c_\mu \ket{\psi_\mu}\ket{0^m} - \sum_\mu c_\mu \ket{\psi_\mu}\ket{\tilde{E}_\mu}\right\| \le  \zeta
\end{align}
where
$\ket{\tilde{E}_\mu}=\ket{\varphi_{\mu,1}\varphi_{\mu,2}\varphi_{\mu,3}\ldots\varphi_{\mu,s}}\otimes\ket{\rest_\mu}$ contains the $s$-bit truncated representation of $\varphi_\mu = E_\mu/E_{\max} = 0.\varphi_{\mu,1}\varphi_{\mu,2}\varphi_{\mu,3}\cdots$,
and $E_{\max}$ is the upper bound on the maximum energy of the target Hamiltonian $H$ used in the construction of $U_\PE^\NN$.
Let 
\begin{align}
    \tilde{E}_\mu = E_{\max}\times (0.\varphi_{\mu,1}\varphi_{\mu,2}\varphi_{\mu,3}\ldots\varphi_{\mu,s}) = E_\mu + O(E_{\max}2^{-s})
\end{align}
be the truncated-approximation to the energy eigenvalue $E_\mu$.

The new spatially sparse circuit $U_\PE^{\text{sparse}}$ now has $t_0= O(R_0N) = O(\poly(n,\zeta^{-1}))$ gates.
However, the computational qubits are moved to places far from their original location due to the swap gates, and the history states are far from a local encoding of the original state.
To address this issue, we construct the following uncomputed circuit with idling
\begin{align}
U_\circuit = (\Id)^L U_{\rm PE}^{\text{sparse}\dag} (\Id)^s U_{\rm PE}^{\text{sparse}},
\end{align}
which we will soon transform into our spatially sparse Hamiltonian.
Note this circuit is automatically spatially sparse since it's a composition of 4 spatially sparse circuits, and the computational states are conveniently moved back to the first column after the uncomputation.
With the addition of $U_{\rm PE}^{\text{sparse} \dag}$ for uncomputing and $s+L$ idling identity gates, the entire circuit gate count is now $T=2t_0+s+L$.
The $s$ identity gates in the middle are used for local ``measurements'' of energy to $s$-bit precision by energy penalty Hamiltonian terms.
The later $L=O((2t_0+s)\|H\|^2/\epsilon^2)=O(\poly(n, \zeta^{-1})/\epsilon^2)$ identity gates are used to ensure $O(\epsilon)$ simulation precision by a trivial encoding as in Lemma~\ref{lem:circuit-idling}.
The history states with respect to $U_\circuit$ and the eigenstates $\{\ket{\psi_\mu}\}$ of $H$ are
\begin{align}
\ket{\eta_\mu} =\frac{1}{\sqrt{T+1}}\sum_{t=0}^T \Big( U_t \cdots U_2 U_1 \ket{\psi_\mu}\ket{0^m} \Big) \ket{1^t 0^{T-t}}.
\end{align}

We now convert the circuit to a Hamiltonian ${H}_\circuit$ using the method described in Lemma~\ref{lem:circuit-ham-simul}, where $H_{out}$ is chosen to be
\begin{align}
\label{eq:Hout-energy-penalty}
H_{out} = (T+1) E_{\max} \sum_{b=1}^s 2^{-b}\ketbra{1}_b^\anc \otimes P^\clock(t_0+b).
\end{align}
We also denote $P^\clock(t)=\ketbra{110}_{t-1,t,t+1}^\clock$, which projects onto legal clock states corresponding to time step $t$.

To show that ${H}_\circuit$ simulates the original Hamiltonian $H$, we first show that $H_{out}$ restricted to the subspace of history states $\L=\spn\{\ket{\eta_\mu}:1\le \mu \le 2^n \}$ can be approximated by the following effective Hamiltonian
\begin{equation}
H_\eff = \sum_\mu E_\mu \ketbra{\eta_\mu}.
\end{equation}
Consider arbitrary states $\ket{\eta}\in \L$.
We write $\ket{\eta} = \sum_{\mu} a_\mu \ket{\eta_\mu}$, and observe
\begin{align}
\braket{\eta| H_{out}|\eta} 
&= E_{\max} \sum_{b=1}^s 2^{-b} \left[\sum_{\nu} a_\nu^* \bra{\psi_\nu}\bra{0^m} \right]U_{\rm PE}^{\text{sparse}\dag} \ketbra{1}_b U_{\rm PE}^\text{sparse} \left[\sum_{\mu} a_\mu  \ket{\psi_\mu}\ket{0^m}\right]
\end{align}
Then using \eqref{eq:PE-approx}, we have
\begin{align}
\braket{\eta |  H_{out} |\eta} 
&= E_{\max} \sum_{b=1}^s 2^{-b} \left[\sum_{\nu} a_\nu^* \bra{\psi_\nu}\bra{\tilde{E}_\nu} + \bra{\zeta}\right] \ketbra{1}_b \left[\sum_{\mu} a_\mu \ket{\psi_\mu}\ket{\tilde{E}_\mu} + \ket{\zeta} \right] 
\end{align}
where $\ket{\zeta}$ is some residual state vector with $\|\ket{\zeta}\|\le \zeta$.
Hence
\begin{align}
\left|\braket{\eta |  H_{out} - H_\eff|\eta}\right| 
&\le \sum_\mu |a_\mu|^2 |\tilde{E}_\mu- E_\mu| + 2s\zeta E_{\max} \le \max_\mu |\tilde{E}_\mu- E_\mu| + 2s\zeta E_{\max} \le (2^{-s} + 2s\zeta ) E_{\max}
\end{align}
We can ensure this is always less than $\epsilon/4$ by choosing for example 
\begin{align}
\label{eq:s-zeta-choice}
    s = \log_2 (8 E_{\max}/\epsilon) = O(\log n + \log \epsilon^{-1})
    \quad
    \text{and} 
    \quad
    \zeta = \epsilon/(16s E_{\max}) = O(1/\poly(n,\epsilon^{-1})).
\end{align}
Hence,
\begin{equation}
\left|\braket{\eta |  H_{out} - H_\eff|\eta}\right| \le \epsilon/4 \quad \forall \ket{\eta} \in \L \quad
\Longrightarrow \quad
\|H_\eff - H_{out}|_{\L} \| \le \epsilon/4
\label{eq:Heff-Hout}
\end{equation}
Furthermore, since we have added $L$ idling gates such that $\|H\otimes \ketbra{\alpha}-H_\eff\|\le \epsilon/4$ for some ancilla state $\ket{\alpha}$ by Lemma~\ref{lem:circuit-idling}, then together with \eqref{eq:Heff-Hout} we have
\begin{equation}
\|H\otimes \ketbra{\alpha} - H_{out}|_{\L} \| \le \epsilon/2.
\end{equation}
Observe that we can rewrite $H\otimes \ketbra{\alpha} = VHV^\dag$, where $V\ket{\psi}=\ket{\psi}\ket{\alpha}$ $\forall \ket\psi\in \mathbb{C}^{2^n}$  is an isometry.
Hence, by Lemma~\ref{lem:circuit-ham-simul}, for any $\eta>0$, the constructed ${H}_\circuit$ simulates $H$ to precision $(\Delta, \eta,\epsilon)$, where $\Delta=O(\epsilon^{-1}\|H_{out}\|^2 + \eta^{-1}\|H_{out}\|) = O(\epsilon^{-1}\|H\|^2 + \eta^{-1}\|H\|)$.
Note that $H_\circuit$ is spatially sparse since $U_{\rm PE}^\text{sparse}$ is spatially sparse.
Since $U_{\rm PE}^\text{sparse}$ contains at most 2-local gates, which means $H_\circuit$ is at most 5-local.
Furthermore, $H_\circuit$ contains $O(T)=O(\poly(n,\zeta^{-1})/\epsilon^2 )=O(\poly(n,\epsilon^{-1}))$ terms (and qubits), with $O(\poly(n,T,\epsilon^{-1},\eta^{-1}, \|H_{out}\|)) = O(\poly(n,\eta^{-1},\epsilon^{-1}))$ interaction energy.

\end{proof}

To finish the proof, we just need to prove Lemma~\ref{lem:circuit-ham-simul} and \ref{lem:circuit-idling}. We start with the proof of Lemma \ref{lem:circuit-ham-simul}.

\begin{proof}[\textbf{Proof of Lemma~\ref{lem:circuit-ham-simul}}]
For a given circuit $U = U_T \cdots U_2 U_1$,
the corresponding circuit-Hamiltonian is
\begin{align}
H_{\circuit} &= H_0 +  H_{out}\\
\text{where} \quad H_0 &= J_{clock} H_{clock} + J_{prop} H_{prop} + J_{in} H_{in}
\end{align}
The role of $H_0$ is to isolate $\L=\spn\{\ket{\eta_\mu}\}$ as its zero-energy groundspace separated by a large spectral gap $2\Delta$ from the rest of the eigenstates.
Then $H_{out}$ is used recover the eigenvalue structrue of $H$ in the subspace $\L$, allowing $H_{\circuit}$ to simulate $H$.

Now we give the explicit form of the circuit-Hamiltonian.
The first part of $H_0$ is
\begin{equation}
\label{eq:Hclock}
H_{clock} = \sum_{t=1}^{T-1}\ketbra{01}^\clock_{t,t+1},
\end{equation}
which sets the legal state configurations in the clock register to be of the form $\ket{t}^\clock \equiv \ket{1^t 0^{T-t}}^\clock$.
Then, we simulate the state propagation under the circuit using
\begin{eqnarray}\label{eq:Hprop}
H_{prop} &=& \sum_{t=1}^T H_{prop,t},\\
\text{where} \quad H_{prop,t} &=& \Id \otimes \ketbra{100}^\clock_{t-1,t,t+1} - U_t\otimes\ketbrat{110}{100}^\clock_{t-1,t,t+1} \nonumber\\
&& - U_t^\dag\otimes \ketbrat{100}{110}^\clock_{t-1,t,t+1} + \Id\otimes\ketbra{110}^\clock_{t-1,t,t+1}
\quad \text{for} \quad 1<t<T, \nonumber\\
H_{prop,1} &=& \Id\otimes \ketbra{00}_{12}^\clock - U_1 \otimes\ketbrat{10}{00}_{12}^\clock - U_1^\dag \otimes \ketbrat{00}{10}_{12}^\clock + \Id\otimes\ketbra{10}_{12}^\clock, \nonumber\\
\text{and} \quad H_{prop,T} &=& \Id\otimes \ketbra{10}_{T-1,T}^\clock - U_T \otimes\ketbrat{11}{10}_{T-1,T}^\clock - U_T^\dag\otimes\ketbrat{10}{11}_{T-1,T}^\clock + \Id\otimes\ketbra{11}_{T-1,T}^\clock.\nonumber
\end{eqnarray}
These terms check the propagation of states from time $t-1$ to $t$ is correct.
Now, we also need to ensure that the input states are valid, i.e. ancilla qudits are in the state $\ket{0^{m}}^\anc$ when $t=0$ (i.e., the clock register is $\ket{0^T}^\clock$).
This can be done using
\begin{gather}
H_{in} = \sum_{i=1}^{m} (\Id - \ketbra{0})_i^\anc\otimes\ketbra{0}^\clock_{t_{\min}(i)},
\\
\text{where}
\quad
t_{\min}(i) = \min \{t: U_t \text{ acts nontrivially on ancilla qudit } i\}.
\nonumber
\end{gather}
In other words, for each ancilla qudit $i$, $H_{in}$ penalizes the ancilla if it's not in the state $\ket{0}$ before it is first used by the $t_{\min}(i)$-th gate.
Note that ${H}_\circuit$ has $O(T)$ terms, each of which is most $(k+3)$-local when $U_t$ are $k$-local.
If $U$ is spatially sparse, then it is easy to see that $H_\circuit$ is also spatially sparse.

Note that $H_0\L=0$.
We then need to lower bound the spectral gap of $H_0$, i.e. $\lambda_1(H_0|_{\L^\perp} )$, where $\lambda_1(H)$ denotes the lowest eigenvalue of $H$.
To that end, let us denote the following subspaces:
\begin{align}
\S_{clock} &= \spn\{ \ket{\psi}\ket{y}\ket{1^t 0^{T-t}}:\ket{\psi}\in \CC^{d^n} \text{ and } \ket{y} \in \CC^{d^m}, 0\le t\le T\},
\\
\S_{prop} &= \spn\{\ket{\eta_\mu, y} \equiv \frac{1}{\sqrt{T+1}}\sum_{t=0}^T \Big(U_t  \cdots U_2 U_1 \ket{\psi}\ket{y}  \Big) \ket{1^t 0^{T-t}}: 1\le \mu \le d^n, 0\le y \le d^{n-1}\}.
\end{align}
Note that $ \L \subset \S_{prop} \subset \S_{clock}$.
Let us denote $\tilde{A}=\A\cap \L^\perp$ for any subspace $\A$.
Note $H_{clock} \S_{clock} = 0$, $H_{prop} \S_{prop} =0$, $H_{in}\L=0$.
We will use the following Projection Lemma~\ref{lem:projection}:
\begin{lemma}[Projection Lemma, adapted from \cite{KKR06}]
 \label{lem:projection}
 Let $H=H_1+H_2$ be the sum of two Hamiltonians operating on some Hilbert space $\S_0=\S\oplus\S^\perp$.
 Assuming that $H_2$ has a zero-energy eigenspace $\S\subseteq \S_0$ so that $H_2\S=0$, and that the minimum eigenvalue $\lambda_1(H_2|_{\S^\perp})\ge J > 2\|H_1\|$, then
 \begin{equation}
 \lambda_1(H_1|_\S) - \frac{\|H_1\|^2}{J-2\|H_1\|} \le \lambda_1 (H) \le \lambda_1 (H_1|_\S).
 \end{equation}
 In particular, if $J \ge K\|H_1\|^2 + 2\|H_1\|=O(K\|H_1\|^2)$, we have
 $
 \lambda_1(H_1|_\S) - \frac{1}{K} \le \lambda_1 (H) \le \lambda_1 (H_1|_\S).
 $
\qedextra
\end{lemma}
Applying the above Lemma successively to $H_0$, we obtain
\begin{align}
\lambda_1(H_0|_{\L^\perp} ) &\ge \lambda_1\left[(J_{prop} H_{prop} + J_{in} H_{in} )|_{\tilde\S_{clock}} \right] - \frac{1}{K}
	\qquad \text{if} \quad J_{clock} = O(K\|J_{prop} H_{prop} + J_{in} H_{in}\|^2) \\
&\ge \lambda_1\left[( J_{in} H_{in})|_{\tilde\S_{prop}} \right] - \frac{2}{K}
	\qquad \text{if} \quad J_{prop}/T^2 = O(K\| J_{in} H_{in}\|^2)
	\label{eq:lambda1}
\end{align}
where we used the fact that $\lambda_1(H_{clock}|_{\S_{clock}^\perp})\ge 1$,
and $\lambda_1(H_{prop}|_{\S_{prop}^\perp}) \ge c/T^2$ for some constant $c$.
We now lower bound \eqref{eq:lambda1}.
Let us denote $\hat{n}=\Id - \ketbra{0}$.
Then within $\S_{clock}$, we can rewrite
\begin{align}
H_{in}|_{\S_{clock}} &= \sum_{i=1}^m \hat{n}_{i}^\anc \otimes \sum_{0\le t \le t_{\min}(i)} \ketbra{t}^\clock = \sum_{t=0}^{\max_i t_{\min}(i)} H_{in,t} \\
\text{where} \quad H_{in,t} &= \sum_{\{i:~t \le t_{\min}(i)\}} \hat{n}_i^\anc \otimes \ketbra{t}^\clock. \nonumber
\end{align}
In particular, $H_{in,t=0}=\sum_{i=1}^m \hat{n}_i^\anc \otimes \ketbra{t=0}$.
Thus, for any $\ket{\eta_\mu,y},\ket{\eta_\nu,y'}\in \tilde{\S}_{prop}$, where necessarily $y,y'>0$, we have
\begin{eqnarray}
\braket{\eta_\nu,y'|H_{in,t=0}|\eta_\mu,y} &=&  \frac{1}{T+1} \bra{\psi_\nu}\bra{y'} H_{in,t=0} \ket{\psi_\mu}\ket{y} \nonumber \\
&=& \frac{1}{T+1} \delta_{\mu\nu} \braket{y'|\sum_{i=1}^{m}\hat{n}_i^\anc |y}=  \frac{1}{T+1} \delta_{\mu\nu}\delta_{y,y'} \times w(y),
\end{eqnarray}
where $w(y)$ is the Hamming weight of $y$ in $d$-ary representation, which is at least 1 for any $y>0$.
Hence, the minimum eigenvalue of $H_{in,t=0}|_{\L^\perp}$ is $1/(T+1)$.
Since $H_{in}$ consists of only positive semi-definite terms, we have \begin{align}
    \lambda_1(H_{in}|_{\tilde{\S}_{prop}}) \ge \lambda_1( H_{in,t=0}|_{\tilde{\S}_{prop}}) \ge 1/(T+1).
\end{align}
Thus, to ensure that $H_0$ has spectral gap $\lambda_1(H_0|_{\L^\perp}) \ge 2\Delta$, we simply choose $J_{in} = O(\Delta(T+1))$, $J_{prop} = O(K T^2 J_{in}^2 m^2)$,
and $J_{clock} = O(KJ_{prop}^2 T^2)=O(\poly(n, T, \Delta))$.

Now that we have shown $H_0$ has $\L$ as its groundspace with spectral gap $2\Delta$,
we are ready to show that $H_\circuit$ $(\Delta,\eta,\epsilon)$-simulates $H$ with only polynomial overhead in energy.
To this end, we use the following result regarding perturbative reductions adapted from Lemma 4 of \cite{BravyiHastingsSim} (also Lemma 35 of \cite{UniversalHamiltonian}):
\begin{lemma}[First-order reduction, adapted from \cite{BravyiHastingsSim}]
\label{lem:BravyiHastingsSim}
Suppose $\tilde{H}=H_0+H_{out}$, defined on Hilbert space $\tilde\H=\L \oplus \L^\perp$ such that $H_0\L=0$ and $\lambda_1(H_0|_{\L^\perp})\ge 2\Delta$.
Suppose $H$ is a Hermitian operator and $V$ is an isometry such that $\| V H V^\dag - H_{out}|_{\L} \| \le \epsilon/2$, then
$\tilde{H}$ is a $(\Delta, \eta,\epsilon)$-simulation of $H$, as long as $\Delta \ge O(\epsilon^{-1}\|H_{out}\|^2 + \eta^{-1}\|H_{out}\|)$, per Definition~\ref{defn:CMPsimul}.
In other words, $\|\tilde{H}_{\le\Delta}  - \tilde{V} H \tilde{V}^\dag\| \le \epsilon$ for some isometry $\tilde{V}$ where $\|\tilde{V}-V\|\le \eta$.
\qedextra
\end{lemma}

Observe we are given in the premise of this Lemma~\ref{lem:circuit-ham-simul} that
\begin{equation}
\|V H V^\dag - H_{out}|_{\L}\| \le \epsilon/2.
\end{equation}
Hence, $H_\circuit = H_0+H_{out}$ simulates $H$ to precision $(\Delta, \eta,\epsilon)$ where $\Delta\ge  O(\epsilon^{-1}\|H_{out}\|^2 + \eta^{-1}\|H_{out}\|)$.
The maximum interaction energy in $H_\circuit$ is $J_{clock} = O(\poly(n, T, \Delta))= O(\poly(n,T,\epsilon^{-1},\eta^{-1}, \|H_{out}\|))$.
This concludes the proof of Lemma~\ref{lem:circuit-ham-simul}.
\end{proof}

We now prove the second Lemma, which shows that in order to ensure the circuit-Hamiltonian simulates the original Hamiltonian with good precision with trivial encoding, we only need to add $O(\poly(n,\epsilon^{-1}))$ ``idling'' identity gates to the end of a polynomial-sized circuit before transforming the circuit back to a Hamiltonian.

\begin{proof}[\textbf{Proof of Lemma~\ref{lem:circuit-idling}}]
Note that we can write
\begin{align}
\ket{\eta_\mu} &= \sqrt{1-\chi^2}\ket{\psi_\mu}\otimes\ket{\alpha} + \chi \ket{\beta_\mu}
\end{align}
where
\begin{align}
\ket{\alpha} &= \frac{1}{\sqrt{L+1}} \ket{0^m}^\anc \otimes \sum_{t=D}^{D+L} \ket{1^t 0^{T-t}}^\clock,  \\
\ket{\beta_\mu} &= \frac{1}{\sqrt{D}}\sum_{t=0}^{D-1}\Big( U_{t}\cdots U_2 U_1 \ket{\psi_\mu}\ket{0^{m}}^\anc \Big) \ket{1^t 0^{T-t}}^\clock, \\
\text{and}\qquad
\chi &= \sqrt{D/(D+L+1)}.
\end{align}
Observe that $\bra{\beta_\mu}(\ket{\psi_\nu}\ket{\alpha}) = 0$ since the clock register are at different times, and
\begin{equation}
\braket{\beta_\mu|\beta_\nu} = \frac{1}{D}\sum_{t=0}^{D-1}\braket{\psi_\mu|\psi_\nu} = \delta_{\mu\nu}.
\end{equation}
Let $P_\anc = \ketbra{\alpha}$. Then
\begin{align}
H_\eff - H \otimes P_\anc &= \sum_\mu \Big[ E_\mu \ketbra{\eta_\mu} - E_\mu \ketbra{\psi_\mu}\otimes \ketbra{\alpha}\Big] \nonumber
\\
&=\bigoplus_\mu 
\begin{pmatrix}
-E_\mu\chi^2 & E_\mu\chi\sqrt{1-\chi^2}  \\
E_\mu\chi\sqrt{1-\chi^2}  & E_\mu\chi^2
\end{pmatrix}
\end{align}
And so
\begin{align}
\|H_\eff - H \otimes P_\anc\| &\le  \chi \max_\mu   E_\mu  \le \chi \|H\|.
\end{align}
To ensure $\|H_\eff - H \otimes P_\anc\| \le \epsilon$,  it's sufficient to choose $L$ so that $\chi \|H\| = \epsilon$.
Plugging in $\chi=\sqrt{D/(D+L+1)}$, we find that it is sufficient to choose $L=O(\frac{D\|H\|^2}{\epsilon^2})$.

\end{proof}

\subsection{Proof of Theorem~\ref{thm:main} -- Strongly Universal Hamiltonian on 2D Square Lattice}
In this subsection, we show how to transform the spatially sparse Hamiltonian constructed previously into a Hamiltonian from a universal family of 2D spin-lattice model, with only polynomial overhead, proving our main Theorem~\ref{thm:main}. This follows from our Proposition~\ref{prop:sparseHam} and the following result from Ref.~\cite{UniversalHamiltonian}:

\begin{lemma}[Essentially Ref.~\cite{UniversalHamiltonian}]
\label{lem:universal2D}
Given any $k$-local $n$-qudit Hamiltonian $H$ that is spatially sparse, we can construct $H'$ from a family of $\S$-Hamiltonian on the 2D square lattice that efficiently $(\Delta,\eta,\epsilon)$-simulates $H$ as long as $\S$ is non-2SLD.
Here, $\Delta = O(\poly(\|H\|,\eta^{-1},\epsilon^{-1})$, and $H'$ has $O(\poly(n,\epsilon^{-1}))$ qubits and interaction energy $O(\poly(\Delta))$.
\end{lemma}

Then our main Theorem is a simple consequence of the above Lemma and our Proposition~\ref{prop:sparseHam}:

\begin{proof}[\textbf{Proof of Theorem~\ref{thm:main}}]
As we showed in Proposition~\ref{prop:sparseHam}, any $O(1)$-local qudit Hamiltonian $H$ with $\|H\|$ can be simulated by a spatially sparse 5-local Hamiltonian $H_{\circuit}$ with $O(\poly(n)/\epsilon^2)$ terms and qubits, and interaction energy at most $O(\poly(n,\eta^{-1},\epsilon^{-1}))$.
By Lemma~\ref{lem:universal2D}, we can simulate $H_\circuit$ by a $\S$-Hamiltonian on the 2D square lattice with polynomial overhead, as long as $\S$ is non-2SLD.
\end{proof}

Although never formally stated, Lemma~\ref{lem:universal2D} is in fact one of the main results of  Ref.~\cite{UniversalHamiltonian}.
Here we provide an outline of its proof for completeness.
\begin{proof}[\textbf{Proof Outline of Lemma~\ref{lem:universal2D}}]
To show this, we use a sequence of reductions originally described in Ref.~\cite{OliveiraTerhal, UniversalHamiltonian},
which together performs the desired transformation.
These reductions are enumerated in the following list of Lemmas:

\begin{lemma}[Lemma 21 of \cite{UniversalHamiltonian}]
\label{lem:qudit-to-qubit}
Given any $k$-local Hamiltonian on $n$ qu$d$its $H$, we can construct a $k\lceil\log_2 d\rceil$-local Hamiltonian $H'$ on $n\lceil\log_2 d\rceil$ qubits that $(\Delta,0,0)$-simulates $H$, for $\Delta\ge \|H\|$. For $d=O(1)$, the construction preserves spatial sparsity, and uses terms of interaction energy $O(\Delta)$.
\end{lemma}
The construction maps each qudit to $\lceil\log_2 d\rceil$ qubits, and uses local terms of strength $\Delta$ to penalize any redundant states among the qubits.
Specifically, consider any isometry $W:\CC^d\to (C^2)^{\otimes \lceil \log_2 d\rceil}$. The construction maps $H$ to $H'= W^{\otimes n}H W^{\dag\otimes n}+ \Delta' \sum_{i=1}^n P_i$, for any $\Delta'> \Delta$, where $P=1-WW^\dag$.
It is easy to see that $H'$ is spatially sparse if $H$ is spatially sparse and $d=O(1)$.

\begin{lemma}[Lemma 22 of  \cite{UniversalHamiltonian}]
\label{lem:complex-to-real}
Given any $k$-local $n$-qubit Hamiltonian $H$, we can construct a real-valued $2k$-local $2n$-qubit Hamiltonian $H'$ that $(\Delta,0,0)$-simulates $H$, for any $\Delta\ge 2\|H\|$. The construction preserves spatial sparsity, and uses terms of interaction energy $O(\Delta)$.
\end{lemma}
The construction adds one additional qubit per original qubit, and map the individual Pauli operators in the following way:
\begin{equation}
\Id \mapsto \Id\otimes \Id, \quad
\sigma_{x,z} \mapsto \Id \otimes \sigma_{x,z}, \quad
\sigma_y \mapsto \sigma_y\otimes \sigma_y.
\end{equation}
For each new pair of qubits $(i,n+i)$, an additional local term $\Delta'(Y_{i}Y_{n+i}+\Id)$ is added, where $\Delta'>\Delta$.
Note the new Hamiltonian is real-valued, and spatially sparse if $H$ is spatially sparse.

\begin{lemma}[Lemma 39 of \cite{UniversalHamiltonian}]
\label{lem:getting-rid-of-Y}
Real-valued $k$-local qubit Hamiltonian $H$ with $M$ terms can be $(\Delta,\eta,\epsilon)$-simulated by real $(k+1)$-local Hamiltonian with $O(M+n)$ qubits and terms, whose Pauli-decomposition contains no $Y$ terms.
The construction preserves spatial sparsity, and uses interaction energy at most $\Delta = O(\poly(\|H\|,\eta^{-1},\epsilon^{-1}))$.
\end{lemma}
The construction here takes any terms in $H$ with (necessarily) even number of $Y$'s, and recreates it with a perturbative gadget involving only $X,Z$ terms and an additional mediator qubit $a$.
Specifically, the gadget performs the following mapping:
\begin{gather}
Y^{\otimes 2m}\otimes A \quad \mapsto \quad \Delta h_0 + \sqrt{\Delta} h_2 \\
\text{where}
\quad
h_0=(1+Z_a)/2=\ketbra{0}_a, \quad
h_2=X_a(X^{\otimes 2m }\otimes \Id + (-1)^{m+1} Z^{\otimes 2m} \otimes A).
\end{gather}
Every term is mapped in parallel with an independent mediator qubit gadget.
Since there are $M$ terms in $H$, we have at most $O(M+n)$ qubits in the end.
The large interaction energy $\Delta = O(\poly(\|H\|,\eta^{-1},\epsilon^{-1}))$ is required to ensure small errors from perturbation.
It is easy to see that if $H$ is spatially sparse, so is the new Hamiltonian.

\begin{lemma}[Theorem 40 of \cite{UniversalHamiltonian}]
\label{lem:k-to-2-local}
Suppose $H$ is any $k$-local qubit Hamiltonian with $M$ terms whose Pauli-decomposition contains no $Y$.
Then $H$ can be simulated by 2-local qubit Hamiltonians with $O(M+n)$ terms and qubits to precision $(\Delta,\eta,\epsilon)$ whose Pauli-decomposition contains no $Y$ terms.
The construction preserves spatial sparsity, and uses interaction energy at most $\Theta(\Delta) = O(\poly(\|H\|, \eta^{-1},\epsilon^{-1}))$ assuming $k=O(1)$.
\end{lemma}
This construction makes use of the subdivision and 3-to-2 local gadgets, which are described in details in Ref.~\cite{OliveiraTerhal}.
For each $k$-local term of the form $A\otimes B$, one can map it to $(\lceil k/2\rceil+1)$-local terms of the form $A\otimes X_w + X_w\otimes B$ using a subdivision gadget that introduces an extra ancilla qubit $w$.
Thus, $O(\log k)$ applications of the subdivision gadget is sufficient to reduce the locality to 3-local.
This is then reduced to 2-local terms using the 3-to-2 local gadget: this converts terms of the form $A\otimes B\otimes C$ to 2-local terms such as $(A-B) X_w$, $C\ketbra{1}_w$, $A B$, and $(A^2+B^2)C$.
In other words, the construction maps each $k$-local term to $O(k)$ 2-local terms mediated by $O(k)$ ancilla qubits.
This mapping clearly preserves spatial sparsity.
The required interaction energy blows up exponentially in $k$; however, since $k=O(1)$, the interaction energy required is at most $O(\poly(\|H\|, \eta^{-1},\epsilon^{-1}))$.

\begin{lemma}[Theorem 41 of \cite{UniversalHamiltonian}]
\label{lem:map-to-XXYYZZ}
Suppose $H$ is a 2-local $n$-qubit Hamiltonian whose Pauli-decomposition contains no $Y$ terms.
Then it can be simulated by a $4n$-qubit $\S$-Hamiltonian to precision $(\Delta,\eta,\epsilon)$, where $\S=\{XX+YY+ZZ\}$ or $\{XX+YY\}$.
The construction preserves spatial sparsity, and uses interaction energy at most $\Theta(\Delta)=O(\poly(\|H\|, \eta^{-1},\epsilon^{-1}))$.
\end{lemma}
Here, the construction uses a perturbative gadget that maps every logical qubit in $H$ to a group of 4 physical qubits that interact only via terms from $\S$.
The 1-local and 2-local interaction on any two logical qubits can be implemented using two-body terms from $\S$ coupling different pairs of physical qubits from the two groups.
Hence, spatial sparsity is preserved by this construction.
The required interaction energy scales as $\Theta(\Delta) = O(\poly(\|H\|, \eta^{-1},\epsilon^{-1}))$.

Finally, we restate the following two results from Ref.~\cite{OliveiraTerhal}:

\begin{lemma}[Lemma 46 of \cite{UniversalHamiltonian}]
\label{lem:map-to-2D-XXYYZZ}
Let $\S_0$ be either $\{XX+YY+ZZ\}$ or $\{XX+YY\}$.
Any spatially sparse $\S_0$-Hamiltonian on $n$ qubits, whose largest interaction energy is $\Lambda_0$, can be simulated by a $\S_0$-Hamiltonian on a 2D square lattice of $\poly(n)$ qubits using interaction energy at most $J_{ij} = O(\poly(n\Lambda_0(1/\epsilon+1/\eta)))$.

\end{lemma}

\begin{lemma}[Theorem 42 of \cite{UniversalHamiltonian}]
\label{lem:map-to-2D-general}
Suppose $\S$ be a set of interactions on 2-qubits that is non-2SLD.
Then given an $\{XX+YY+ZZ\}$- or $\{XX+YY\}$-Hamiltonian on the 2D square lattice, we can simulate it with an $\S$-Hamiltonian on the 2D square lattice.
\end{lemma}

In what follows, we denote $\S_0$ as either $\{XX+YY+ZZ\}$ or $\{XX+YY\}$.
For any $\S$ that is non-2SLD, we map $H_\circuit$ to an $\S$-Hamiltonian on the 2D square lattice in the following sequence:
\begin{enumerate}
\item By Lemma~\ref{lem:qudit-to-qubit}, we can simulate $H_\circuit$ with $H_1$ that is spatially sparse, $O(1)$-local on $O(\poly(n)/\epsilon^2)$ qubits and interaction energy at most $O(\|H_\circuit\|)=O(\poly(n,\eta^-1,\epsilon^-1))$.
\item By Lemma~\ref{lem:complex-to-real}, we can simulate $H_1$ with $H_2$ that is spatially sparse, real-valued, and $O(1)$-local, with only polynomial overhead in qubit-number of interaction energy.
\item By Lemma~\ref{lem:getting-rid-of-Y}, we can simulate $H_2$ with $H_3$ that is spatially sparse and $O(1)$-local, and contains no $Y$ terms in Pauli-decomposition, with only polynomial overhead in qubit-number of interaction energy.
\item By Lemma~\ref{lem:k-to-2-local}, we can simulate $H_3$ with $H_4$ that is spatially sparse and 2-local, contains no $Y$ terms in Pauli-decomposition, with polynomial overhead.
\item By Lemma~\ref{lem:map-to-XXYYZZ}, we can simulate $H_4$ with $H_5$ that is a spatially sparse $\S_0$-Hamiltonian, with polynomial overhead.
\item By Lemma~\ref{lem:map-to-2D-XXYYZZ}, we can simulate $H_5$ by $H_6$ an $\S_0$-Hamiltonian on a 2D square lattice, with polynomial overhead.
\item By Lemma~\ref{lem:map-to-2D-general}, we can simulate $H_6$ by the broader class of $\S$-Hamiltonian on the 2D square lattice, for any $\S$ that is non-2SLD, with polynomial overhead.
\end{enumerate}
Since every step of the above sequence of reductions only incurs a polynomial overhead in the number of qubits and the strength of interactions, we have shown that any $O(1)$-local, polynomial-sized qudit Hamiltonians can be efficiently simulated by an $\S$-Hamiltonian with polynomial qubits and interaction energy, assuming $\S$ is non-2SLD.
\end{proof}


\section{Proof that 1D nearest-neighbor Hamiltonians are strongly universal}
\label{supp4}
Here we give our proof of Theorem~\ref{thm:1D-universal}, whose statement we reproduce below for convenience:
\begin{thm}
\label{thm:1D-universal}
There is a strongly universal family of 1D Hamiltonians consisting of nearest-neighbor interactions acting on a line of 8-dimensional particles.
\end{thm}
The proof is based heavily on the framework in Ref.~\cite{Hallgren}.
We will only try to provide a succinct and somewhat self-contained description of the most important elements of the construction here.
For the full technical details of the construction, we encourage the reader to also examine Section 3 and 4 of Ref.~\cite{Hallgren}.

\subsection{Preliminaries}
We first describe how the computation is encoded within a line of 8-dimensional particles.

\begin{defn}[1D-encoded $L$-idling history state]
Consider any quantum circuit $U$ consisting of $R=O(\poly(n))$ rounds of 1-qubit or nearest-neighbor 2-qubit gates on a line of $n$ qubits.
This can be further converted to an \emph{encoded circuit} $\tilde{U}$ with nearest-neighbor gates acting on a line $2nR+L$ qu$d$its ($d=8$), implicitly arranged in $R$ blocks of $2n$ qudits, followed by $L$ qudits for idling.
The Hilbert space of each qudit is $\H_8=\blnk\oplus\lmove\oplus \insi \oplus \dead \oplus \qubit \oplus \gate$, where $\qubit$ and $\gate$ are 2-dimensional subspaces designed to hold a qubit state, and the rest are 1-dimensional subspaces.
For a given input state of the form $\ket{\psi_\mu}\ket{0^m}\in\CC^{2^n}$ on the original $n$ qubits, this is encoded as
\begin{align} \label{initialconfiguration}
\ket{\gamma_0^\mu}
=
\overbrace{
	\bdry \underbrace{\gate \insi \parity \qubit \insi \parity \mgdots \parity \qubit \insi \parity \qubit \blnk }_\text{\rm the first block of length $2n$}
	\bdry \blnk \blnk \parity \blnk \blnk \parity \mgdots  \mgdots 
	}^{R \textnormal{ blocks}}
	\clkbdry \underbrace{\blnk\blnk\blnk \cdots \blnk}_{L \textnormal{ idling qudits}}
\end{align}
where the odd qudits in the first block of length $2n$ encodes the input state $\ket{\psi_\mu}\ket{0^m}$.
Here, the symbols $\bdry$, $\parity$ and $\clkbdry$ are simply boundary markers in space that help us identify the role of each particle and do not indicate anything about the internal state of particles.
In particular, the symbol $\clkbdry$ marks a special boundary that separates the computational part of the line and the idling part.

The encoded circuit $\tilde{U}$ acts on $\ket{\gamma_0^\mu}$ with $R$ rounds of computation, each corresponding to applying a round of gates  from $U'$ to the currently active block of qudits, and then moving the block $2n$ positions to the right.
This entails a total of $K=(R-1)(3n^2+2n-1)+2n$ steps of nearest-neighbor gates that map configuration to configuration, according to the transition rule outlined in Table 1 of Ref.~\cite{Hallgren}.
This is then followed by $L$ steps of ``idling'' where the $L$ rightmost qudits perform a trivial counting operation.
To facilitate the idling, we add the following transition rules.
%
%
\begin{itemize}
\item[($\alpha$)] $\gate \clkbdry  \blnk \longleftrightarrow \qubit \clkbdry \dead $ unmarks the active qubit $\gate$ once it hits the special boundary $\clkbdry$. The $\blnk$ changes to $\dead$ so as to signal that the idling is supposed to the start.
\item[($\beta$)] $\dead\blnk \longleftrightarrow \dead\dead$ for any location to the right of the special boundary $\clkbdry$.
\end{itemize}
This results in a history of $K+L+1$ configurations on the $2nR$ qudits $\{\ket{\gamma_t^\mu}\}_{t=0}^{K+L}$ which is pairwise orthogonal: $\braket{\gamma_t^\mu|\gamma_{t'}^\mu}=\delta_{tt'}$. Finally, we define the \emph{1D-encoded $L$-idling history state} with respect to $U$ and $\ket{\psi_\mu}\ket{0^m}$ as the following superposition state:
\begin{align}
\ket{\eta_\mu} = \frac{1}{\sqrt{K+L+1}} \sum_{t=0}^{K+L} \ket{\gamma_t^\mu}.
\end{align}
\end{defn}

To help visualize the computational history according to the transition rules in Ref.~\cite{Hallgren}, we remark that the configuration after all $R$ rounds of computation comes to halt is
\begin{align}
\ket{\gamma_K^\mu} = 
\bdry \dead^{2n(R-1)} \bdry \dead \qubit \parity \insi
\qubit \parity \mgdots \parity \insi \qubit \parity \insi \gate
\clkbdry \blnk^{\otimes L}.
\end{align}
All ensuing configurations are of the form (for $1\le \ell \le L$):
\begin{align}
\ket{\gamma_{K+\ell}^\mu} = 
\bdry \dead^{2n(R-1)} \bdry \dead \qubit \parity \insi
\qubit \parity \mgdots \parity \insi \qubit \parity \insi \qubit
\clkbdry \dead^{\otimes \ell} \blnk^{\otimes (L-\ell)}.
\end{align}

\begin{lemma}[1D circuit-Hamiltonian \cite{Hallgren}]
\label{lem:1DcircuitHam}
Given a circuit $U$ consisting of $\poly(n)$ 1- or 2-qubit gates on $n$ qubits.
We can construct a Hamiltonian on $O(\poly(n))+L$ qu$d$its, $d=8$, with only nearest-neighbor interaction of the form
\begin{align}
H_{\rm hist} = J_{\rm in} H_{\rm in} + J_{\rm prop} H_{\rm prop} + J_{\rm pen} H_{\rm pen}
\end{align}
such that the 1D-encoded $L$-idling history states with respect to $U$ and $\ket{\psi_\mu}\ket{0^m}$ are of zero-energy, and all other states have energy $\ge 1$.
The interaction energy of nearest-neighbor terms in $H_{\rm hist}$ are $J_{\rm in}, J_{\rm prop}, J_{\rm pen}= O(\poly(n))$.
\end{lemma}
\begin{proof}
The construction is essentially the same as described in Section 4 of Ref.~\cite{Hallgren}, except for a few small changes:
\paragraph{Changes to legal configurations and penalty Hamiltonian}---
 To the right of the special boundary $\clkbdry$, only $\dead\dead$, $\dead\blnk$ and $\blnk\blnk$ are allowed configurations in these locations. This can be addressed by tweaking the penalty Hamiltonian $H_{\rm pen}$ to penalize all configurations using the term $\ketbra{XY}_{i,i+1}$ where $XY\in \H_8^{\otimes 2} \setminus \{\dead\dead, \dead\blnk, \blnk\blnk \}$ for these locations.

\paragraph{Changes to propagation Hamiltonian}---
We need to incorporate the two new rules $(\alpha)$ and $(\beta)$ added above. The Rule $(\alpha)$ is similar to Rule 4a $\gate\bdry\blnk \longleftrightarrow \qubit \bdry \lmove$ from Table~1 of Ref.~\cite{Hallgren}. Since there's only a unique location where Rule $(\alpha)$ applies, at the special boundary, we can simply use the following propagation Hamiltonian for that pair of sites
\begin{align}
H_{\textrm{prop},i}^{(\alpha)} = \ketbra{\gate\clkbdry\blnk}_{i,i+1} + \ketbra{\qubit\clkbdry\dead}_{i,i+1} - \ketbrat{\qubit\clkbdry\dead}{\gate\clkbdry\blnk}_{i,i+1} - \ketbrat{\gate\clkbdry\blnk}{\qubit\clkbdry\dead}_{i,i+1}
\end{align}
where $i=2{n}R$ for this special pair of sites.

Now for Rule $(\beta)$. Note this is the only propagation rule that is applicable in the region to the right of the special boundary $\clkbdry$. Thus, we only need the following propagation Hamiltonian for $i>2{n}R$.
\begin{align}
H_{\textrm{prop},i}^{(\beta)} = \ketbra{\dead\blnk}_{i,i+1} + \ketbra{\dead\dead}_{i,i+1} - \ketbrat{\dead\dead}{\dead\blnk}_{i,i+1} - \ketbrat{\dead\blnk}{\dead\dead}_{i,i+1}
\end{align}

We note there can be mis-timed transitions from $H_{\textrm{prop},i}^{(\alpha)}$, e.g.
\begin{align}
\insi \qubit\clkbdry\dead\dead
\quad \longrightarrow \quad
- \insi \gate\clkbdry\blnk\dead
\end{align}
 and $H_{\textrm{prop},i}^{(\alpha)}$, e.g.,
\begin{align}
\dead\dead\dead \quad \longrightarrow \quad -\dead\blnk\dead
\end{align}
However, they will all result in energy penalty from $H_{\rm pen}$, because they have illegal configuration $\blnk\dead$ that is locally detectable.

\paragraph{Proof that the Hamiltonian has the 1D-encoded history states as the only ground states}---
This is essentially given in Section 5 and 6 of \cite{Hallgren}.
\end{proof}

\subsection{Proof of Theorem~\ref{thm:1D-universal}}

\begin{proof}[\textbf{Proof of Theorem~\ref{thm:1D-universal}}]
As in the Proof of Theorem~\ref{thm:main}, we can always convert any input qudit Hamiltonian to a qubit Hamiltonian by encoding each qu$d$it in a group of $\lceil\log_2 d\rceil$-qudits with polynomial overhead (assuming $d=O(1)$).
Hence, we will take the input $H$ as an $O(1)$-local $n$-qubit Hamiltonian.
We write $H=\sum_\mu E_\mu \ketbra{\psi_\mu}$ in its eigenbasis, with $0\le E_\mu \le E_{\max}$.
By Proposition~\ref{prop:PEcircuit}, we can construct a circuit $U_\PE^\NN$ consisting on $O(\poly (n,\zeta^{-1}))$ 1- or 2-qubit nearest-neighbor gates on $n+m$ qubits, $m=O(\poly(n))$, such that its action on 
any normalized state $\sum_\mu c_\mu \ket{\psi_\mu}$ can be described as
\begin{align}
\left\| U_\PE^\NN \sum_\mu c_\mu \ket{\psi_\mu}\ket{0^m} - \sum_\mu c_\mu \ket{\psi_\mu}\ket{\tilde{E}_\mu} \ket{\rest_\mu}\right\| \le  \zeta
\end{align}
where $\ket{\tilde{E}_\mu}=\ket{\varphi_{\mu,1}\varphi_{\mu,2}\varphi_{\mu,3}\ldots\varphi_{\mu,s}}$ is the $s$-bit string representation of $\varphi_\mu = E_\mu/E_{\max} = 0.\varphi_{\mu,1}\varphi_{\mu,2}\varphi_{\mu,3}\cdots$.
From now on we will denote $\tilde{n}=n+m$ as the number of qubits in the phase estimation circuit.

We then consider the identity circuit $U=(U_\PE^\NN)^\dag \Id^{n+m} U_\PE^\NN=\Id$ that corresponds to running $U_\PE^\NN$, applying identity to every qubit, and then uncomputing.
Let $R_{\PE}$ be the number of rounds of nearest-neighbor gates in $U_{\PE}$, and $R=2R_\PE + 1$ be the total number of rounds.
Let $K=(R-1)(3\tilde{n}^2+2\tilde{n}-1)+2\tilde{n}$ be the total number of steps when applying $U$.
By Lemma~\ref{lem:1DcircuitHam}, we can construct a 1D nearest-neighbor Hamiltonian $H_{\rm hist}$ on $\poly(n)+L$ particles with $\poly(n)$ interaction energy such that all the 1D-encoded $L$-idling history states with respect to $U$ and $\ket{\psi_\mu}\ket{0^m}$ are the only zero-energy states, and all other states have energy at least $\Delta$.

We claim that the following 1D Hamiltonian $(\Delta, \eta,\epsilon)$-simulates $H$
\begin{align}
\tilde{H}_{\rm 1D} = 2\Delta H_{\rm hist} + H_{\rm out}.
\end{align}
To describe $H_{\rm out}$, we start by considering the configuration of all $2\tilde{n}R+L$ particles after the $R_\PE$-th round of gates corresponding to the end of applying $U_\PE^\NN$.
This looks like
\begin{align}
\cdots \dead\dead \bdry \gate \insi \parity \qubit \insi \parity \cdots \qubit \insi \parity \qubit \blnk \bdry \blnk \blnk \cdots \blnk \clkbdry \blnk \cdots
\end{align}
Note the next round of gates corresponds to $\Id^{n+m}$, i.e., applying the identity gate to every qubit.
Without loss of generality, we assume the first $s$ active particles corresponds ($\zeta$-approximately) to the ancilla with $\ket{\tilde{E}_\mu}$ on them.
Notice that every active qudit will at one point transition into their appropriate state in the $\gate$ subspace, with identity gate applied.
Then, the appropriate $H_{\rm out}$ is
\begin{align}
H_{out} = (K+L+1) E_{\max} \sum_{b=1}^s 2^{-b} \ketbra{\gate^{(1)}}_{2\tilde{n}R_{\PE} + 2b-1}
\end{align}
where $\ket{\gate^{(1)}}$ corresponds to the $\ket{1}$ qubit state in the $\gate$ subspace.

To prove that $\tilde{H}_{\rm 1D}$ simulates $H$ to the desired precision, we first show that $H_{\rm out}$ restricted to the set of 1D-encoded history states $\L =\spn\{\ket{\eta_\mu}\}$ can be approximated by the following effective Hamiltonian:
\begin{align}
H_\eff = \sum_\mu E_\mu \ketbra{\eta_\mu}
\end{align}
Consider arbitrary states $\ket{\eta}\in \L$.
We write $\ket{\eta} = \sum_{\mu} a_\mu \ket{\eta_\mu}$, and observe
\begin{align}
\braket{\eta| H_{out}|\eta}  &= (K+L+1)E_{\max}\sum_{b=1}^s 2^{-b} \sum_{\nu, \mu} a_\nu^*a_\mu \bra{\eta_\nu} \left( \ketbra{\gate^{(1)}}_{2\tilde{n}R_{\PE} + 2b-1} \right) \ket{ \eta_{\mu}} \nonumber \\
&= E_{\max}\sum_{b=1}^s 2^{-b} \sum_{\nu, \mu}\sum_{t,t'=0}^{K} a_\nu^*a_\mu \bra{\gamma_t^\nu} \left( \ketbra{\gate^{(1)}}_{2\tilde{n}R_{\PE} + 2b-1} \right) \ket{\gamma_{t'}^{\mu}}
\end{align}
Note the $2\tilde{n}R_{\PE}+2b-1$-th particle is only in the $\gate$ subspace at one point in time among the $K+1$ steps, regardless of possible input state. Table 2 of \cite{Hallgren} provides a clear illustration for the reason.
Let's call that time $t_b$, and write
\begin{align}
\braket{\eta|H_{out}|\eta} 
&= E_{\max}\sum_{b=1}^s 2^{-b} \sum_{\nu, \mu} a_\nu^*a_\mu \bra{\gamma_{t_b}^\nu} \left( \ketbra{\gate^{(1)}}_{2\tilde{n}R_{\PE} + 2b-1} \right) \ket{\gamma_{t_b}^{\mu}} 
\end{align}
Note the configuration $\sum_\mu a_\mu \ket{\gamma_{t_b}^\mu}$ is the encoded version of $U_{\PE}^\NN \sum_\mu a_\mu\ket{\psi_\mu} \ket{0^m}$, which is $\zeta$-close to the encoded version of $\sum_\mu a_\mu \ket{\psi_\mu}\ket{\tilde{E}_\mu}\ket{\rest_\mu}$.
Hence
\begin{align}
\braket{\eta|H_{out}|\eta} &= E_{\max}\sum_{b=1}^s 2^{-b} \sum_{\nu, \mu} a_\nu^*a_\mu \varphi_{\mu,b} \braket{\psi_\nu|\psi_\mu}\braket{\rest_\nu|\rest_\mu} +  \Theta(\zeta E_{\max}) \nonumber \\
&= \sum_\mu |a_\mu|^2\sum_{b=1}^s 2^{-b}\varphi_{\mu,b} E_{\max} +  \Theta(\zeta E_{\max}) \nonumber \\
&= \sum_\mu |a_\mu|^2 \tilde{E}_\mu +  \Theta(\zeta E_{\max}).
\end{align}
where we have denoted $\tilde{E}_\mu = E_{\max} \tilde{\varphi}_\mu = E_{\max}(0.\varphi_{\mu,1}\varphi_{\mu,2}\cdots\varphi_{\mu,s})$ as the $s$-bit representation of energy eigenvalue $E_\mu$ of $H$.
Then
\begin{align}
\left|\braket{\eta |  H_{out} - H_\eff|\eta}\right| 
&\le \sum_\mu |a_\mu|^2 |\tilde{E}_\mu- E_\mu| + \Theta(s\zeta E_{\max})  \le [2^{-s} + \Theta(\zeta) ] E_{\max}
\end{align}
By choosing $s = \Theta(\log E_{\max}/\epsilon)=\Theta(\log n + \log \epsilon^{-1})$ and $\zeta = \Theta(\epsilon/E_{\max}) = \Theta(1/\poly(n,\epsilon^{-1}))$, just like we did in Eq.~\eqref{eq:s-zeta-choice}, we can ensure
\begin{equation}
\left|\braket{\eta |  H_{out} - H_\eff|\eta}\right| \le \epsilon/4 \quad \forall \ket{\eta} \in \L \quad
\Longrightarrow \quad
\|H_\eff - H_{out}|_{\L} \| \le \epsilon/4
\label{eq:1D-Heff-Hout}
\end{equation}

Furthermore, looking at the 1D-encoded $L$-idling history state more carefully, it looks like
\begin{align}
\ket{\eta_\mu} &=\frac{1}{\sqrt{K+L+1}}\sum_{t=0}^{K+L}\ket{\gamma_t^\mu} = \sqrt{1-\chi^2}\ket{\alpha_\mu} + \chi\ket{\beta_\mu}
\end{align}
where
\begin{align}
\chi &= \sqrt{\frac{K+1}{K+L+1}}
\end{align}
and
\begin{align}
\ket{\beta_\mu} &= \frac{1}{\sqrt{K+1}}\sum_{t=0}^K \ket{\gamma_t^\mu} \\
\ket{\alpha_\mu} &= \frac{1}{\sqrt{L}}\sum_{t=1}^{L} \ket{\gamma_{K+t}^\mu} = 
\bdry \dead^{2\tilde{n}(R-1)} \bdry
\dead \qubit \parity \insi \qubit \parity \mgdots \parity \insi \qubit \parity \insi \qubit
\clkbdry 
\frac{1}{\sqrt{L}}\sum_{\ell=1}^L \dead^{\otimes \ell} \blnk^{\otimes (L-\ell)} \nonumber \\
&= (V\ket{\psi_\mu}) \otimes \ket{\alpha}
\end{align}
where $V\ket{\psi_\mu}$ is simply the input state $\ket{\psi_\mu}$ stored in the qubit subspace of a subset of the $\tilde{n}$ qudits marked with $\qubit$ in the $R$-th block, $\ket{\alpha}$ is the state of the remaining ancilla. Therefore,
\begin{align}
\ket{\eta_\mu} = \sqrt{1-\chi^2}(V\ket{\psi_\mu})\otimes \ket{\alpha} + \chi \ket{\beta_\mu}.
\end{align}
By choosing $L=O(K/\epsilon^2)$, we can ensure that $\|\ketbra{\eta_\mu} - V\ketbra{\psi_\mu}V^\dag \otimes \ketbra{\alpha}\|\le O(\epsilon)$.
With the same argument in Lemma~\ref{lem:circuit-idling}, we can show that
\begin{align}
\|H_\eff - VHV^\dag \otimes \ketbra{\alpha}\| \le \epsilon/4
\end{align}
Therefore, together with Eq.~\eqref{eq:1D-Heff-Hout} we have
\begin{align}
\left\|VHV^\dag \otimes \ketbra{\alpha} - H_{out}|_{\L}\right \|\le \epsilon/2
\end{align}

To finish proving our Theorem~\ref{thm:1D-universal}, we again use Lemma~\ref{lem:BravyiHastingsSim} which we restate below for the reader's convenience:
{
\renewcommand{\thelemma}{\ref{lem:BravyiHastingsSim}}
\begin{lemma}[First-order reduction, adapted from \cite{BravyiHastingsSim}]
Suppose $\tilde{H}=H_0+H_1$, defined on Hilbert space $\tilde\H=\L \oplus \L^\perp$ such that $H_0\L=0$ and $\lambda_1(H_0|_{\L^\perp})\ge 2\Delta$.
Suppose $H$ is a Hermitian operator and $V$ is an isometry such that $\| V H V^\dag - H_1|_{\L} \| \le \epsilon/2$, then
$\tilde{H}$ $(\Delta, \eta,\epsilon)$-simulates $H$, as long as $\Delta \ge O(\epsilon^{-1}\|H_1\|^2 + \eta^{-1}\|H_1\|)$, per Definition~\ref{defn:CMPsimul}.
In other words, $\|\tilde{H}_{\le\Delta}  - \tilde{V} H \tilde{V}^\dag\| \le \epsilon$ for some isometry $\tilde{V}$ where $\|\tilde{V}-V\|\le \eta$.
\end{lemma}
\addtocounter{lemma}{-1}
}
We apply the above Lemma with $H_0=2\Delta H_{\rm hist}$ and $H_1 = H_{out}$.
Thus, $\tilde{H}_{\rm 1D}=H_0+H_1$ simulates $H$ to precision $(\Delta, \eta, \epsilon)$ by choosing $\Delta \ge O((\epsilon^{-1}+\eta^{-1})\poly(n,\|H\|))$. Note that $\tilde{H}_{\rm 1D}$ contains $O(\poly(n,\epsilon^{-1}))$ nearest-neighbor terms, with interaction energy $O(\poly(n,\eta^{-1},\epsilon^{-1},\Delta))$, which proves our Theorem.
\end{proof}

\end{document}